\makeatletter\def\@nil{}\makeatother

\documentclass{aa}  
\usepackage{graphicx}
\usepackage{booktabs}
\usepackage{txfonts}
\usepackage{xcolor}
\usepackage{float}
\usepackage{caption}

\usepackage{blindtext}

\usepackage{booktabs}

\usepackage{multirow}

\usepackage{threeparttable} 

\newcommand{\Jyb}{Jy beam$^{-1}$}
\newcommand{\Jya}{Jy arcsec$^{-2}$}

\begin{document}

   \title{Diffuse Radio Emission from Galaxy Clusters in the LOFAR Two-metre Sky Survey Deep Fields}


   \author{E. Osinga
          \inst{1}
          \and 
          R. J. van Weeren\inst{1}
          \and
          J. Boxelaar\inst{1}
          \and 
          G. Brunetti\inst{2}
          \and
          A. Botteon\inst{1}
          \and 
          M. Br\"uggen\inst{3}
          \and
          T. W. Shimwell\inst{4,1}
          \and 
          A. Bonafede\inst{5,2}
          \and
          P.N. Best\inst{6}
          \and 
          M. Bonato\inst{2,7,8}
          \and 
          R. Cassano\inst{2}
          \and 
          F. Gastaldello\inst{9}
          \and 
          G. di Gennaro\inst{1}
          \and 
          M. J. Hardcastle\inst{10}
          \and 
          S. Mandal\inst{1}
          \and 
          M. Rossetti\inst{9}
          \and 
          H. J. A. R\"ottgering\inst{1}
          \and
          J. Sabater\inst{6}
          }

    \institute{Leiden Observatory, Leiden University, PO Box 9513, NL-2300 RA Leiden, The Netherlands \email{osinga@strw.leidenuniv.nl}
    \and 
    Istituto Nazionale di Astrofisica, Istituto di Radioastronomia Via P Gobetti 101, 40129 Bologna, Italy
    \and 
    Hamburg Observatory, University of Hamburg, Gojenbergsweg 112, 21029 Hamburg, Germany
    \and
    ASTRON, the Netherlands Institute for Radio Astronomy, Postbus 2, NL-7990 AA Dwingeloo, The Netherlands
    \and
    DIFA - Universit\'a di Bologna, via Gobetti 93/2, I-40129 Bologna, Italy
    \and 
    SUPA, Institute for Astronomy, Royal Observatory, Blackford Hill, Edinburgh, EH9 3HJ, UK
    \and
    Italian ALMA Regional Centre, Via Gobetti 101, I-40129, Bologna, Italy
    \and
    INAF-Osservatorio Astronomico di Padova, Vicolo dell'Osservatorio 5, I-35122, Padova, Italy
    \and 
    INAF - IASF Milano, via A. Corti 12, 20133 Milano, Italy
    \and 
    Centre for Astrophysics Research, University of Hertfordshire, College Lane, Hatfield AL10 9AB, UK
    %
    }

   \date{Received X; accepted Y}

 
  \abstract
    {Low-frequency radio observations are revealing an increasing number of diffuse synchrotron sources from galaxy clusters, dominantly in the form of radio halos or radio relics. The existence of this diffuse synchrotron emission indicates the presence of relativistic particles and magnetic fields. It is still an open question what mechanisms exactly are responsible for the population of relativistic electrons driving this synchrotron emission. The LOFAR Two-metre Sky Survey Deep Fields offer a unique view of this problem. Reaching noise levels below 30 $\mu$Jy/beam, these are the deepest images made at the low frequency of 144 MHz. This paper presents a search for diffuse emission in galaxy clusters in the first data release of the LOFAR Deep Fields. We detect a new high-redshift radio halo with a flux density of $8.9 \pm 1.0$ mJy and corresponding luminosity of $P_{144\mathrm{MHz}}=(3.6 \pm 0.6)\times10^{25}$ W Hz$^{-1}$ in an X-ray detected cluster at $z=0.77$ with a mass estimate of $M_{500} = 3.3_{-1.7}^{+1.1} \times 10^{14} M_\odot.$ 
    Deep upper limits are placed on clusters with non-detections. 
    We compare the results to the correlation between halo luminosity and cluster mass derived for radio halos found in the literature. This study is one of few to find diffuse emission in low mass ($M_{500} < 5\times10^{14} M_\odot$) systems and shows that deep low-frequency observations of galaxy clusters are fundamental for opening up a new part of parameter space in the study of non-thermal phenomena in galaxy clusters.
    }


   \maketitle
%

\section{Introduction}\label{sec:introduction}
Galaxy clusters are the largest virialised conglomerations of baryons and dark matter in the Universe as well as the densest parts of the large-scale matter structure of the Universe. An increasing number of galaxy clusters are revealing diffuse synchrotron radio emission, which indicates the presence of magnetic fields and a pool of relativistic electrons in the intra-cluster medium (ICM) \citep{WeerenReview}. The properties and origin of the pool of relativistic electrons are still not fully clear \citep{BrunettiJones}, and neither are the exact properties of the magnetic fields of galaxy clusters \citep{2018SSRv..214..122D}. 

The diffuse radio emission in merging galaxy clusters has been broadly classified into two main classes: radio halos and radio relics \citep{2012A&ARv..20...54F,WeerenReview}. Radio halos are diffuse radio structures that roughly follow the thermal ICM  distribution as observed by X-ray observations. Radio relics, also called radio shocks, are elongated and polarized structures found in the outskirts of galaxy clusters that are tracing merger-induced shock waves \citep{BrunettiJones,2012SSRv..166..187B}.

The currently favoured model for radio halos is the turbulent re-acceleration model, which poses that merger-induced turbulence (re-)accelerates cosmic-ray electrons which produce the radio halo \citep[e.g.,][]{2007MNRAS.378..245B,2011ApJ...726...17P,2015ApJ...800...60M}. The turbulent re-acceleration model is supported by observations that show that radio halos are generally found in merging systems \citep[e.g.,][]{2010A&A...517A..10C,Cassano2013,2013MNRAS.436..275W,2015A&A...579A..92K,2015A&A...580A..97C,2017ApJ...843L..29E}. A possible contribution may come from the hadronic model, which states that relativistic electrons are products of hadronic collisions between relativistic protons and thermal ions \citep[e.g.,][]{1999APh....12..169B,2000A&A...362..151D}. 
However, upper limits to gamma-ray emission expected from the decay products, in particular upper limits on the Coma cluster \citep[e.g.,][]{2011ApJ...728...53J,2014MNRAS.440..663Z,2012MNRAS.426..956B,2017MNRAS.472.1506B}, and the very steep spectra observed in a fraction of radio halos \citep[e.g.,][]{2008Natur.455..944B,Wilber_2017} rule out a dominant contribution from this channel, although a scenario where secondaries are re-accelerated by turbulence is not excluded \citep[e.g.,][]{2011MNRAS.410..127B,2017MNRAS.465.4800P,2017MNRAS.472.1506B}.

Radio halos are more commonly found in higher-mass clusters, owing to the known scaling relation between the radio power and host cluster X-ray luminosity or mass \citep{2000ApJ...544..686L,Cassano2013,2019MNRAS.487.4775B}. This scaling relation was found to exhibit a bi-modal behaviour, with merging systems lying on the correlation and with more relaxed systems generally being less luminous or undetected in the radio band at a level significantly below the correlation \citep[e.g.,][]{Cassano2013,2015A&A...580A..97C}. This behaviour corroborates the idea that the kinetic energy dissipated during merger events powers radio halos. 

Some exceptions to the scaling relation and merger connection have been found. There are a few cases of over-luminous radio halos (i.e., halos found in low X-ray luminosity clusters) \citep[e.g.,][]{2009A&A...507.1257G,2011A&A...530L...5G}, although with only a few detections, the classification of these sources remains uncertain.
Radio halos have also been found to be present in (semi-) relaxed clusters \citep{2014MNRAS.444L..44B,2017MNRAS.466..996S,2019A&A...622A..24S}, suggesting that minor mergers in massive clusters might also have the potential to dissipate enough energy to power cluster-scale emission, although again, these are only a few examples.

Most radio halos observed at GHz frequencies have spectral indices slightly lower than $\alpha=-1$ (where $S_\nu \propto \nu^\alpha$) \citep{2009A&A...507.1257G,2012A&ARv..20...54F}. In a number of cases ultra-steep ($\alpha<-1.6$) spectrum radio halos (USSRH) have been observed \citep[e.g.,][]{2009ApJ...699.1288D,2013A&A...551A.141M,Wilber_2017}. The turbulent re-acceleration model predicts that less energetic mergers, often associated with lower mass systems, could generate halos with lower synchrotron break frequencies ($<1$GHz) \citep{2010A&A...517A..10C}. Observing radio halos close to the break frequency leads to finding steeper spectrum halos. Because USSRHs are expected to be discovered at low frequencies, and to be associated mainly to low mass clusters, the correlation between the radio halo luminosity at 120 MHz and the X-ray luminosity of the cluster is predicted to be steeper and more scattered than at higher radio frequencies \citep{2010A&A...517A..10C}. 

There are still many open questions relating to the origin and formation of radio halos. Due to the higher occurrence rate and radio luminosity of halos with increasing cluster mass \citep{Cassano2013,2015A&A...580A..97C}, most of the understanding has been built on studies of relatively massive ($>5\times10^{14} M_\odot$) galaxy clusters. However, it is important to study radio halos in low mass systems to understand their origin. Only a few radio halos have been detected below cluster masses of $5\times10^{14} M_\odot$, with the lowest mass cluster
being A3562 \citep{2003A&A...402..913V} at $2.44^{+0.21}_{-0.24}\times10^{14} M_\odot$ \citep[see ][for a recent compilation of halos from the literature]{2019MNRAS.487.4775B}.

The fact that the turbulent re-acceleration model predicts that an increasing fraction of halos in lower mass clusters should have a steep spectrum implies that lower mass systems should be more easily detected at lower frequencies \citep[e.g.,][]{2010A&A...517A..10C,BrunettiJones}. Furthermore, less massive clusters have a smaller turbulent energy budget, which implies that the effect of turbulent re-acceleration may become less dominant at lower cluster masses. Consequently, a possible transition from turbulent halos to halos powered by hadronic interactions is predicted \citep[e.g.,][]{2012A&A...548A.100C,BrunettiJones}. The transition depends on the amount of cosmic ray protons available in galaxy clusters, which is still not understood.

The LOFAR Deep Fields \citep{Tasse2020,Sabater2020,Kondapally2020,Duncan2020}
are a set of deep LOFAR observations on three fields which have high-quality multi-wavelength ancillary data available. These fields provide a unique opportunity to study radio halos in the low-mass and low luminosity regime due to the low-frequency and large depth of the observations. 
This relatively unexplored regime can elucidate mechanisms of halo formation in low mass clusters that exhibit lower levels of turbulent motions.
In this paper, we present a search for diffuse emission associated with galaxy clusters in the LOFAR Deep Fields. 
Throughout, we assume a flat $\Lambda$CDM cosmology with $H_0=70$kms$^{-1}$Mpc$^{-1}$, $\Omega_m$=0.3 and $\Omega_\Lambda=0.7$. We define the spectral nature of the radio emission as $S_\nu \propto \nu^\alpha$ where $S_\nu$ is the measured flux density at the frequency $\nu$ and $\alpha$ is the spectral index.


\section{Data}\label{sec:data}
The LOFAR surveys key science project aims to survey the Northern sky at 120-168 MHz at several depth tiers with the LOFAR High Band Antenna. The wide survey aims to reach a sensitivity of 100 $\mu$\Jyb over the entire northern sky \citep{2017A&A...598A.104S,2019A&A...622A...1S}, while the Deep Fields are a set of deeper images of a few selected fields. This paper makes use of the first data release of the LOFAR Two Metre Sky Survey (LoTSS) Deep Fields \citep{Tasse2020,Sabater2020,Kondapally2020,Duncan2020}, which currently consists of three fields with a wealth of multi-wavelength data available: the European Large-Area ISO Survey-North 1 (ELAIS-N1; \citealt{2000MNRAS.316..749O}), Bo\"otes \citep{1999ASPC..191..111J} and the Lockman Hole \citep{1986ApJ...302..432L}, which cover a combined area of $>50$ deg$^2$. The final aim of the LoTSS Deep Fields is to reach noise levels of $10-15\mu$Jy beam$^{-1}$ \citep{Tasse2020} near the pointing centre. 

In the first data release, the Lockman Hole and Bo\"otes field were observed for 80 and 112 hours, reaching noise levels in the centre of the fields of $\sim$ 22 and 32 $\mu$Jy beam$^{-1}$ respectively. The observations and data reduction process of these two fields are described in detail by \citet{Tasse2020}. The ELAIS-N1 field was observed for 170 hours, reaching noise levels of $\sim$ 20 $\mu$\Jyb. This field required a custom data reduction strategy due to a different observing setup and bandwidth coverage, which is detailed by \citet{Sabater2020}.

\begin{table*}[tbh]
\centering
\caption{Sample of sources extracted from the LOFAR Deep Fields. The mass, redshift and $R_{500}$ are obtained from the PSZ2 catalogue \citep{2016A&A...594A..27P} or from the MCXC catalogue \citep{2011A&A...534A.109P} if the source was not present in the former catalogue, unless otherwise noted. For the WHL clusters $M_{500}$ was estimated from the richness, as detailed in Section \ref{sec:WHL}. Mass uncertainties are not available for MCXC clusters.}
\begin{threeparttable}[b]

\begin{tabular}{@{}llllll@{}}
\toprule
Source Name & Field & Redshift & $M_{500}$ ($10^{14} M_\odot$) & $R_{500}$ (Mpc) & Radio classification \\ \midrule
MCXC J1033.8+5703 & Lockman & 0.0463 & 0.128 & 0.35 & No detection \\
MCXC J1036.1+5713 & Lockman & 0.7699 & 3.25 & 0.78 & Halo\\
MCXC J1053.3+5720 & Lockman & 0.34 & 0.487 & 0.49 & No detection \\
PSZ2 G147.88+53.24 & Lockman & 0.60 & 6.47 $\pm$ 0.60 & 1.06 & Halo\\
PSZ2 G149.22+54.18 & Lockman & 0.1369 & $5.87_{-0.22}^{+0.23}$ & 1.22 & Halo \\
SpARCS1049+56 & Lockman	& 1.71\tnote{1} & 2.52 $\pm$ 0.86\tnote{2} & 0.51 & AGN\\
SDSSC4-3094 & Lockman	& 0.04632\tnote{3} & & & AGN\\
PSZRX G084.01+46.28 & ELAIS-N1 & 0.0675 &  $1.37_{-0.36}^{+0.33}$ & 0.77 & No detection\\
PSZ2 G084.69+42.28 & ELAIS-N1 & 0.13 & $2.70_{-0.26}^{+0.27}$ & 0.94 & Uncertain\\
WHL J160439.5+543139 & ELAIS-N1 & 0.2655 & $2.95 \pm 0.50$ & 0.93 & Detection uncertain\\
WHL J161135.9+541635 & ELAIS-N1 & 0.3407 & $3.40 \pm 0.58$ & 0.94 & No detection \\
WHL J161420.1+544254 & ELAIS-N1 & 0.3273 & $2.85 \pm0.48$ & 0.89 & Detection uncertain\\ \bottomrule
\end{tabular}%

\begin{tablenotes} 
     \item[1] \citep{2015ApJ...809..173W}
     \item[2] Derived from $M_{200}$ given in \citet{2020ApJ...893...10F}, see Section \ref{sec:SpARCS}.
     \item[3] \citep{2005AJ....130..968M}
   \end{tablenotes}
  \end{threeparttable}%
\label{tab:mysample}
\end{table*}

\section{Methods}

\begin{figure*}[tbh]
    \centering
    \includegraphics[width=1.0\linewidth]{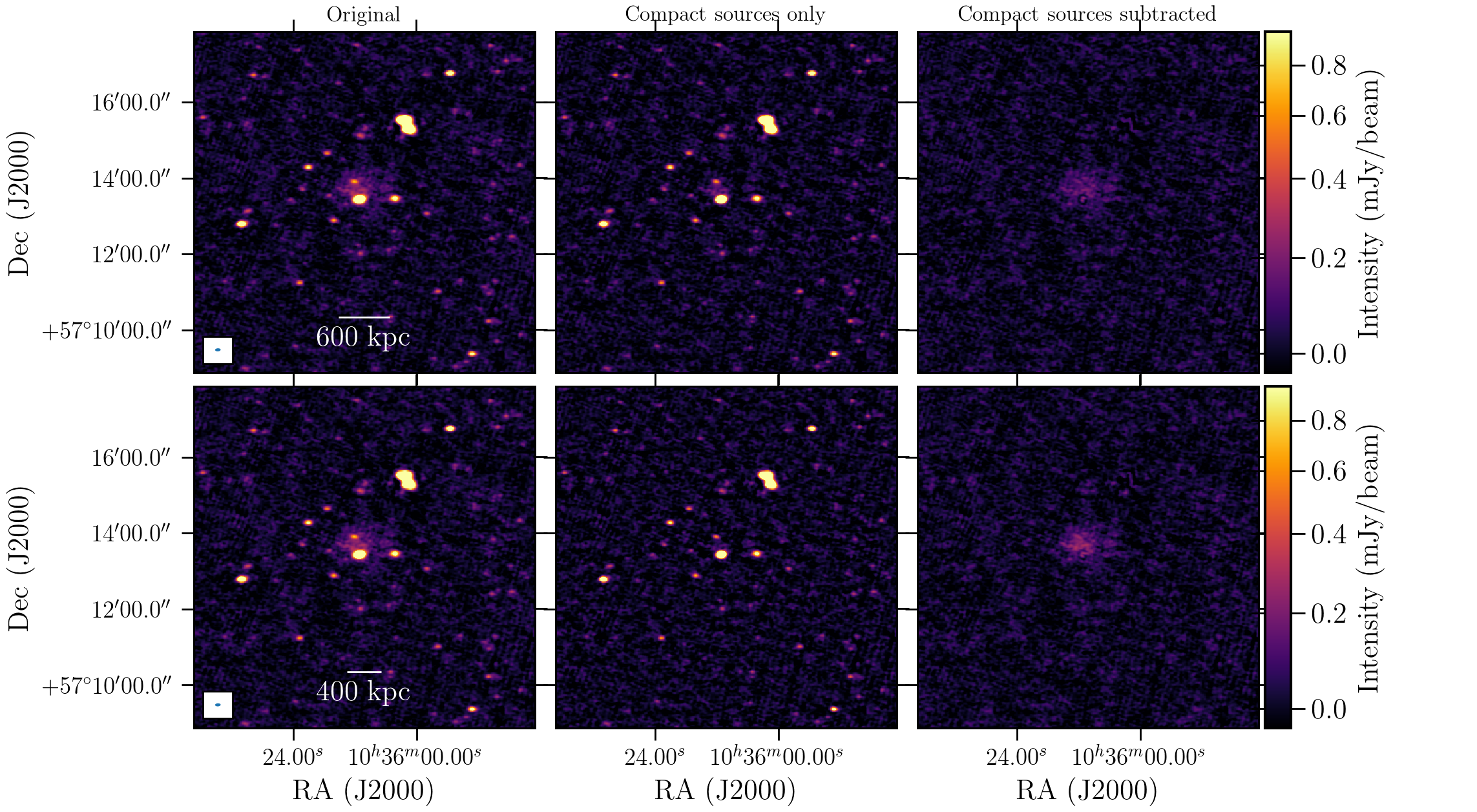}
    \caption{The compact source subtraction process for the cluster MCXCJ1036.1+5713. The left column shows the original image, the middle column the image containing only the compact emission, and the right column shows the final compact source subtracted image. All images are made with the same parameters, except the $uv$-cut for the central image. The top compact source only image was made with a $uv$-cut of 2547$\lambda$ (i.e., 600 kpc at the cluster redshift) while the bottom image was made with a $uv$-cut of 3820$\lambda$ (i.e., 400 kpc). The restoring beam size is 9''$\times$5''.}
    \label{fig:MCXCJ036}
\end{figure*}

We identified all clusters within 2.5 degrees of the pointing centre in the ELAIS-N1 and Lockman Hole fields that were present in the second \textit{Planck} catalogue of Sunyaev-Zel'dovich detected sources \citep[PSZ2;][]{2016A&A...594A..27P}, the Meta-Catalogue of X-ray detected Clusters of galaxies \citep[MCXC;][]{2011A&A...534A.109P} or the Combined Planck-RASS catalogue of X-ray-SZ clusters \citep[ComPRASS;][]{2019A&A...626A...7T}. Seven clusters in the aforementioned catalogues are present in the deep fields, of which the details are given in Table \ref{tab:mysample}. We also checked the optically (SDSS data) selected cluster catalogue WHL \citep{2012ApJS..199...34W} for clusters showing signs of diffuse emission and visually identified three more clusters that show hints of diffuse emission, although these are more likely to be AGN-related extended emission. Finally, we also add the SpARCS1049+56 cluster to our sample, which was identified by \citet{2015ApJ...809..173W} to be a very high redshift ($z=1.71$) cluster in the Lockman Hole field. As a fortuitous bonus, the cluster SDSSC4-3094 identified in the Sloan Digital Sky Survey \citep{2005AJ....130..968M} at $z=0.04632\pm0.00083$ lies in the same region of the sky as SpARCS1049+56 and is therefore added to the sample. Details on the total of 12 clusters are given in Table \ref{tab:mysample}, seven and five of which are in Lockman and ELAIS-N1 respectively. The Bo\"otes deep field observations do not overlap with any clusters from the PSZ2, MCXC or ComPRASS catalogues. No clear diffuse emission from cluster objects was picked up from visual identification of the field, including the 12 spectroscopically confirmed clusters at $z>1$ found by \citet{2008ApJ...684..905E}.

\subsection{Target extraction and imaging}
Once identified, we follow an `extract and subtract' procedure to optimise the sensitivity of the deep images to diffuse emission in the direction of the cluster by allowing for easy re-imaging. First, we make small ($\sim 0.3^\circ \times0.3^\circ$) boxes around the identified targets. The $uv$-data corresponding to this box is extracted from the full dataset with the following method. A direction-dependent calibrated model, from the pipeline described in \citet{Tasse2020,Sabater2020}, of all components outside the boxed region is subtracted from the model data.
This leaves visibilities that contain only sources in the boxed region. We then phase-shift to the center of the extracted region, average the data in time and frequency to reduce the size and perform 7 rounds of direction dependent self calibration with the DDF pipeline\footnote{https://github.com/mhardcastle/ddf-pipeline} \citep{2014arXiv1410.8706T,2015MNRAS.449.2668S,2018A&A...611A..87T} to improve the quality of the extracted image compared to that in the original deep field maps. In the original deep field maps, the facets used for direction dependent calibration are larger as the distance to the pointing center increases, causing more calibration errors related to the assumed constant beam model and ionosphere over a single facet. This extraction procedure mitigates these errors by manually defining a smaller sub-region than the original facet around the target of interest.
The primary beam correction on these extracted and self-calibrated visibilities is done by multiplication with a constant factor of the primary beam response at the centre of the extracted region, which is a good assumption, since the extraction region is much smaller than the size of the LOFAR primary beam (full width at half maximum $\sim 2.5 ^\circ$).
The details of the extraction process are described in \citet{2020arXiv201102387V} and the method has been used with various other LOFAR observations \citep[e.g.,][]{2019MNRAS.488.3416H,2020A&A...634A...4M,2020ApJ...897...93B,2020MNRAS.499L..11B}.

To properly disentangle the extended diffuse emission from compact sources, compact sources are subtracted. This is done as follows: first, an image of the compact sources only is made by ignoring short baselines that are sensitive to extended emission. The inner $uv$-cut is initially calculated such that it corresponds to emission of a certain largest linear physical size at the cluster redshift, based on the mass of the cluster. As an example, for the low mass system MCXCJ1036.1+5713 we found that the $uv$-cut of 2547$\lambda$ (i.e., 600 kpc at $z=0.76991$) was too small to properly exclude all diffuse emission. A $uv$-cut of 3820$\lambda$ (i.e., 400 kpc) shows better separation of diffuse emission and compact sources, as shown in Figure \ref{fig:MCXCJ036}. As higher mass clusters often have larger radio halos, it makes sense to have lower $uv$-cuts (in kpc) with lower cluster mass. 

The clean component model of the compact image is subtracted from the visibilities of the extracted dataset, leaving only the visibilities corresponding to the diffuse extended emission. This emission is imaged with a Gaussian taper corresponding to 50 kpc at the cluster redshift, using multi-scale clean, with \texttt{WSClean} (version 2.7.3) \citep{10.1093/mnras/stu1368,2017MNRAS.471..301O} to properly deconvolve the diffuse emission. The complete compact source subtraction process is illustrated in Figure \ref{fig:MCXCJ036} for the cluster MCXCJ1036.1+5713 as an example. 

\subsection{Measuring radio halo properties}\label{sec:MeasProperties}
To measure the properties of the diffuse emission, we fit the radio halos with an exponential profile. This has a few advantages over manually defining the halo region. Commonly, the radio halo flux density is measured by integrating the surface brightness over an area bounded by isophotes (e.g., 3$\sigma$ contours). However, this causes the resulting flux density to be dependent on the sensitivity of the observations. It is more rigorous to fit the halos with a profile and analytically integrate that profile up to a certain radius. It has been shown that exponential profiles can provide characteristic scales relatively independent of the sensitivity of the radio observations \citep{radial_profile}. In this work we consider the simplest, spherically symmetric, exponential profile for most of the halos, which has been found to be representative of radio halos \citep{radial_profile}, although in some cases observations of radio halos have shown strong deviations from spherical symmetry \citep[e.g.,][]{2016ApJ...818..204V,2018ApJ...856..162W}.
The surface brightness model is given by
    \begin{equation}\label{eq:circ}
        I(\vec{r}) = I_0 e^{-\vec{r}/r_e},
    \end{equation}
where $I_0$ and $r_e$ are the central surface brightness and the $e$-folding radius, respectively. To compare the $e$-folding radii of the halos to the radii of the halos that are normally quoted in the literature ($r_H$), we assume $r_H/r_e=2.6$, as was found by \citet{10.1093/mnras/stx1475} for $8$ clusters with measured $r_H$ within $3\sigma$ isophotes and fitted $r_e$.

    The presented fitting of Equation \eqref{eq:circ} and halo radio flux density estimations were done with a newly developed algorithm\footnote{\url{https://github.com/JortBox/Halo-FDCA}}. The algorithm is described in detail by \citet{Boxelaar2020}, and we briefly explain it here. 
    The fitting algorithm is based on fitting methods first presented by \citet{radial_profile}. The difference here is that profiles are fitted to a two-dimensional image directly rather than to a radially averaged one-dimensional data array. This allows fitting of a non-circular model as well, although for simplicity we assume a circular model in this work. Theoretically, one could fit both a circular and a non-circular model and compare a goodness of fit statistic (e.g., reduced $\chi^2$) of both models to determine which model is a better fit. However, the determination of the morphology of the diffuse emission is beyond the goal of this paper and requires high signal-to-noise data to determine statistically significant differences in the goodness of fit statistic.
    
    The total flux density $S$ of the fitted radio emission is obtained by integrating Eq. \eqref{eq:circ}. The analytical expression for the total flux density is $S=2\pi I_0r_e^2\left( 1-e^{-d}(d+1) \right)$, where $d$ denotes the radius (in $e$-folding radii) up to which is being integrated. 
    Here we choose to integrate up to $2.6r_e$, following \citet{10.1093/mnras/stx1475}. For comparison, integrating up to $2.6r_e$ results in a total flux density that is $73\%$ of the flux density found when integrating the model to infinity. 
    
    The best-fit estimates for the peak surface brightness and $e$-folding radius are found through Bayesian inference and maximum likelihood estimation. To sample the likelihood function, we use a Monte Carlo Markov Chain, implemented within the \texttt{emcee} module \citep{emcee}. This method allows us to find the full posterior distribution for the model parameters. 
    Given observed data $V(\vec{r}_i)$ (which represents the radio surface brightness at position $\vec{r}_i$) and fit parameter vector $\vec{\theta}=(I_0,r_e)$, we assume that all the compact source subtracted images can be expressed as $V(\vec{r}_i) = I(\vec{r}_i)+\epsilon_i$ where $I(\vec{r})$ is defined in Eq. \eqref{eq:circ} and the underlying noise $\epsilon_i$ is independent and identically distributed as $\mathcal{N}(0,\sigma^2_{\mathrm{rms}})$. Independence of individual pixels is assured through re-gridding the images such that the pixel area approximately equals the beam area, while preserving the total flux. 
    The probability density function $f(\vec{r}_i;\vec{\theta})$ for an observation then reads
    \begin{equation}
     f(\vec{r}_i;\theta)=\frac{1}{\sqrt{2\pi}\sigma_{\mathrm{rms}}}\text{exp}\left(-\frac{(V(\vec{r}_i) - I(\vec{r}_i;\theta))^2}{2\sigma^2_{\mathrm{rms}}}\right).
    \end{equation} 
    This results in a log likelihood function which is given by 
    \begin{equation}
     \ln{\mathcal{L}(\theta)}= -n\ln{\sqrt{2\pi}\sigma_{\text{rms}}}-\frac{1}{2\sigma^2_{\text{rms}}}\sum_{i=1}^n (V(\vec{r}_i) - I(\vec{r}_i;\theta))^2,
    \end{equation}
    where the sum is taken over the $n$ re-gridded pixels. Maximising the log-likelihood function for $\theta$ allows us to find the best-fit model parameter vector $\hat{\theta}$.


The uncertainty of the total flux density of the halos $f_H$ is calculated by adding the uncertainty due to map noise (i.e., the uncertainty on best-fit parameters $\sigma_{\mathrm{fit}}$), the absolute flux density scale $\delta_{\mathrm{cal}}$ and compact source subtraction $\sigma_{\mathrm{sub}}$ in quadrature \citep[cf.][]{Cassano2013}.
\begin{equation}\label{eq:errorprop}
    \sigma_{f_H} = \sqrt{ (\delta_{\mathrm{cal}} f_H)^2 + \sigma_{\mathrm{fit}}^2 + \sigma_{\mathrm{sub}}^2}.
\end{equation}
The uncertainty on the best-fit parameters $\sigma_{\mathrm{fit}}$ is given by the 16th and 84th percentile of the converged MCMC chain (i.e., $1\sigma$) and we assume a 10\% error on the absolute flux scale of the LOFAR images $\delta_{\mathrm{cal}}$ \citep{Sabater2020} and a 1\% error on the compact source subtraction process $\sigma_{\mathrm{sub}}$. The latter error is calculated as 1\% of the flux contained in the compact sources only image within $2.6 r_e$ of the center of the fitted halo. This 1\% error is consistent with measuring the residual flux in the compact source subtracted images at the location of bright compact sources. 

For determining the upper limits in the case of non-detections, we use a similar method to that of \citet{10.1093/mnras/stx1475}, which injects mock halos into the visibility data \citep[see also][]{2007ApJ...670L...5B,2008A&A...484..327V}. We inject mock halos following the exponential profile in Eq. \eqref{eq:circ}. Following \citet{10.1093/mnras/stx1475}, we add power spectrum fluctuations of the form $P(\Lambda) \propto \Lambda^n$, where $\Lambda$ is the spatial scale, to account for surface brightness fluctuations observed in real radio halos. We set $\Lambda$ between 10-250 kpc and $n=11/3$ \citep{2005A&A...430L...5G,2006A&A...460..425G,10.1093/mnras/stx1475}.

The initial value of $I_0$ and $r_e$ are chosen such that the expected radio power of the halo follows the $P_{1.4\mathrm{GHz}}-M_{500}$ correlation by \citet{Cassano2013}. Specifically, we first calculate $P_{1.4\mathrm{GHz}}$ from the cluster mass, then set $r_e$ according to the correlation between $P_{1.4\mathrm{GHz}}-r_e$ given by \citet{radial_profile} and finally scale $I_0$ such that the exponential model integrates up to the expected radio power of the halo.
The resulting model is injected (i.e., Fourier transformed and added) into the visibility data at a location close to the cluster but absent of contaminating radio sources. The data is then cleaned and imaged in the same way as the original image.  
We define the halo as detected if the 3$\sigma_{\mathrm{rms}}$ contours cover at least 3 beams. Provided the halo is detected, the $I_0$ is gradually lowered by steps of $\sigma_{\mathrm{rms}}$ to find a more stringent upper limit on the radio power. Conversely, if the halo is not detected, the $I_0$ is gradually increased until it is detected. We inject halos close to the clusters instead of on the center of the clusters to avoid being biased low on the upper limits. Residual emission from point sources or an undetected halo near the cluster center might otherwise contribute to flux measurement of the injected halos. Off cluster injection does assume, however, that there are no calibration artefacts due to bright sources in the cluster.

\begin{figure*}[tbh]
    \centering
    \includegraphics[width=1.0\linewidth]{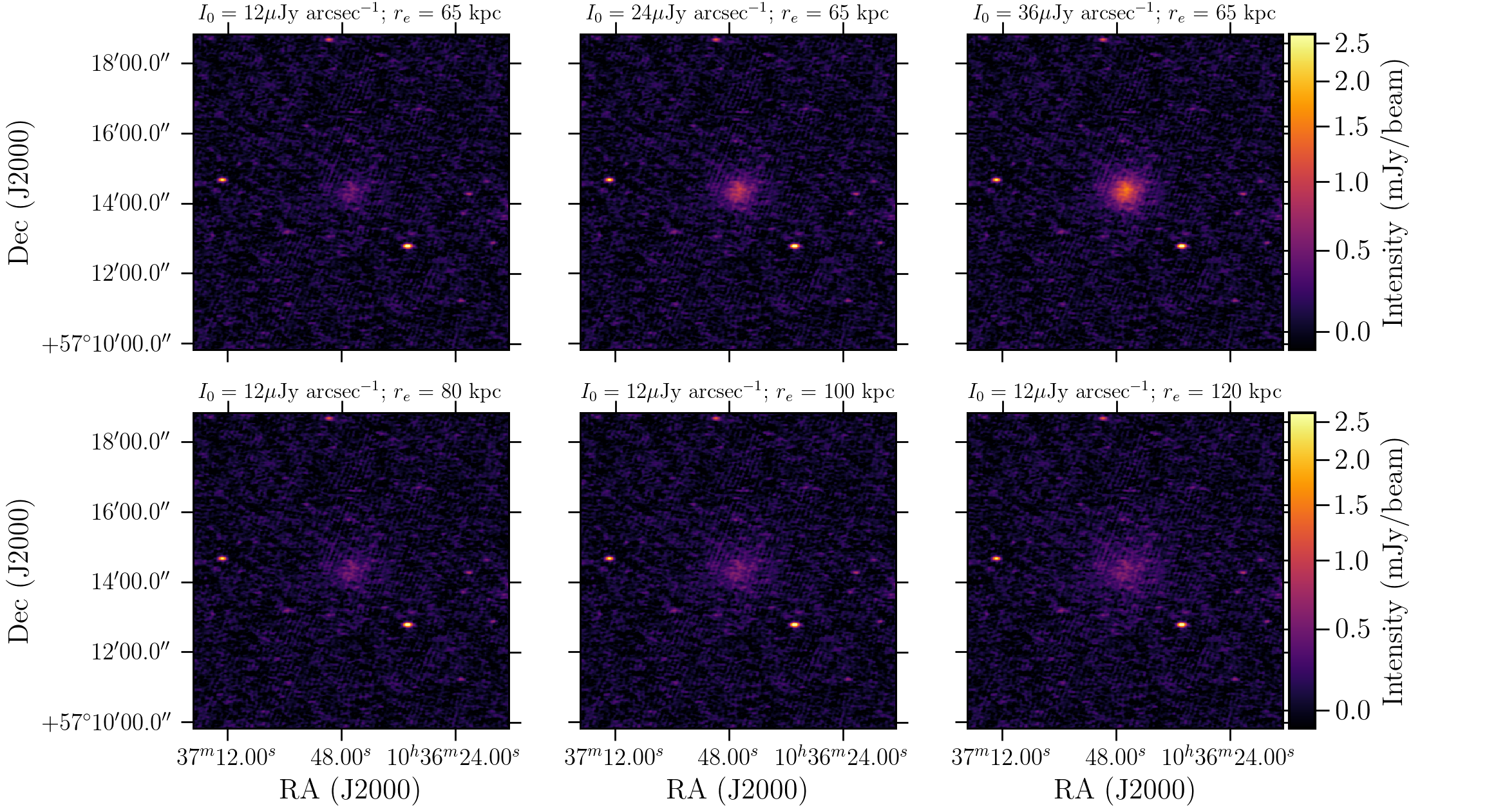}
    \caption{Six mock halos injected into a single 8 hour LOFAR observation of the Lockman Hole field. The different properties of the halos are given in the title figures.}
    \label{fig:inject}
\end{figure*}

\section{Verification on simulated halos}\label{sec:inject}
When determining the properties of diffuse radio emission, it is important to not only keep track of the statistical uncertainties, but to also consider additional sources of error. We test in this section two main effects. The first is the effect of the limited $uv$ coverage of radio telescopes, particularly at shorter baselines, which may cause resolving out some diffuse emission. The second is the point source subtraction process, which may also erroneously subtract some diffuse emission, depending on the $uv$-cut used.

To test the fitting procedure, the sensitivity of the LOFAR observations to different scales of emission and the point source subtraction process, we inject mock halos with different $I_0$ and $r_e$ into a single LOFAR observation ($\sim$ 8 hours of data) of the Lockman Hole field. The full observations are not used for this test due to the computational intensity of the imaging and point-source subtraction process on the full dataset. The local rms at the region of injection is around 100 $\mu$\Jyb. We assume a redshift of $z=0.20$ for the conversion of the $e$-folding radius to angular size. We then compare the injected properties with the properties derived from fitting. Six different halos have been injected into the data, which are shown in Fig. \ref{fig:inject}.

\begin{table*}[]
\centering
\caption{Results of fitting exponential profiles to mock halos of different properties. The injected flux density and resulting flux density are defined as the flux fitted within $2.6r_e$. Note that the uncertainty in the flux density here does not include an absolute flux scale uncertainty, since the injected halo flux densities would change accordingly. }
\label{tab:injectable}
\resizebox{\textwidth}{!}{%
\begin{tabular}{@{}llllll@{}}
\toprule
$I_0$ inject [$\mu$Jy arcsec$^{-2}$] & $r_e$ inject [kpc] & $S_{\nu}$ inject [mJy] & $I_0$ observed [$\mu$Jy arcsec$^{-2}$] & $r_e$ observed [kpc] & $S_{\nu}$ observed [mJy] \\ \midrule
12.0 & 65  & 20.9 & $8.6_{-0.9}^{+1.0}$  & $82_{-8}^{+8}$  & $23.7 \pm 3.5$ \\
24.0 & 65  & 41.9 & $17.3_{-0.9}^{+0.9}$ & $79_{-3}^{+3}$  & $44.5 \pm 5.0$ \\
36.0 & 65  & 62.8 & $26.3_{-0.8}^{+0.9}$ & $77_{-2}^{+2}$  & $63.8 \pm 6.8$ \\
12.0 & 80  & 31.7 & $10.2_{-0.8}^{+0.8}$ & $94_{-6}^{+7}$  & $36.7 \pm 4.7$ \\
12.0 & 100 & 49.6 & $10.0_{-0.7}^{+0.6}$ & $118_{-6}^{+7}$ & $57.9 \pm 6.9$ \\
12.0 & 120 & 71.4 & $10.2_{-0.6}^{+0.5}$ & $140_{-5}^{+7}$ & $83.2 \pm 9.4$ \\ \bottomrule
\end{tabular}%
}
\end{table*}

We subtract point sources by employing a $uv$-cut of 200 kpc, corresponding to $3443\lambda$. The compact source subtracted images are then fitted following the procedure outlined in Section \ref{sec:MeasProperties}. The resulting best-fit parameters and injected parameters are given in Table \ref{tab:injectable}. We find that we generally recover the correct flux density within the 68\% uncertainty, although we are biased slightly higher than the injected flux density. This is because some of the central brightness structure of the mock halos is subtracted out by the compact source subtraction process. This causes generally underestimated $I_0$ and overestimated $r_e$, which also causes generally slightly overestimated $S_\nu$ of about $10\%$, because the integrated flux density scales with $r_e^2$. This bias is important to keep in mind throughout the rest of the paper. The test does show that LOFAR is sensitive to the extended emission of halos following an exponential profile, since we are not resolving out a significant amount of flux. This is in line with what for example \citet{2018MNRAS.478.2218H} and \citet{2020arXiv200604808B} have found for the injection of larger halos into LOFAR observations. The full observations are about a factor of $\sim 3$ deeper than the single pointing used here, so we do not expect significant difference from these results for halos with a central surface brightness down as low as $I_0 \sim 4 \mu$Jy arcsec$^{-2}$.


\section{Results}\label{sec:Selection}
Here we report the results of the fitting procedure for each cluster. Unless otherwise stated, we have performed the fitting on the compact source subtracted images tapered to a resolution corresponding to 50 kpc at the cluster redshift. To calculate the radio luminosity, we assume a spectral index of $\alpha=-1.5\pm0.2$ for clusters where spectral index estimates are not available. We choose this range to cover the typical spectra of halos, including steep-spectrum halos \citep{WeerenReview}. The azimuthally averaged surface brightness profiles and corner plots of the MCMC chain can be found in Appendix II.

\subsection{PSZ2G147.88+53.24}

\begin{figure*}[tbh]
    \centering
    \includegraphics[width=1.0\textwidth]{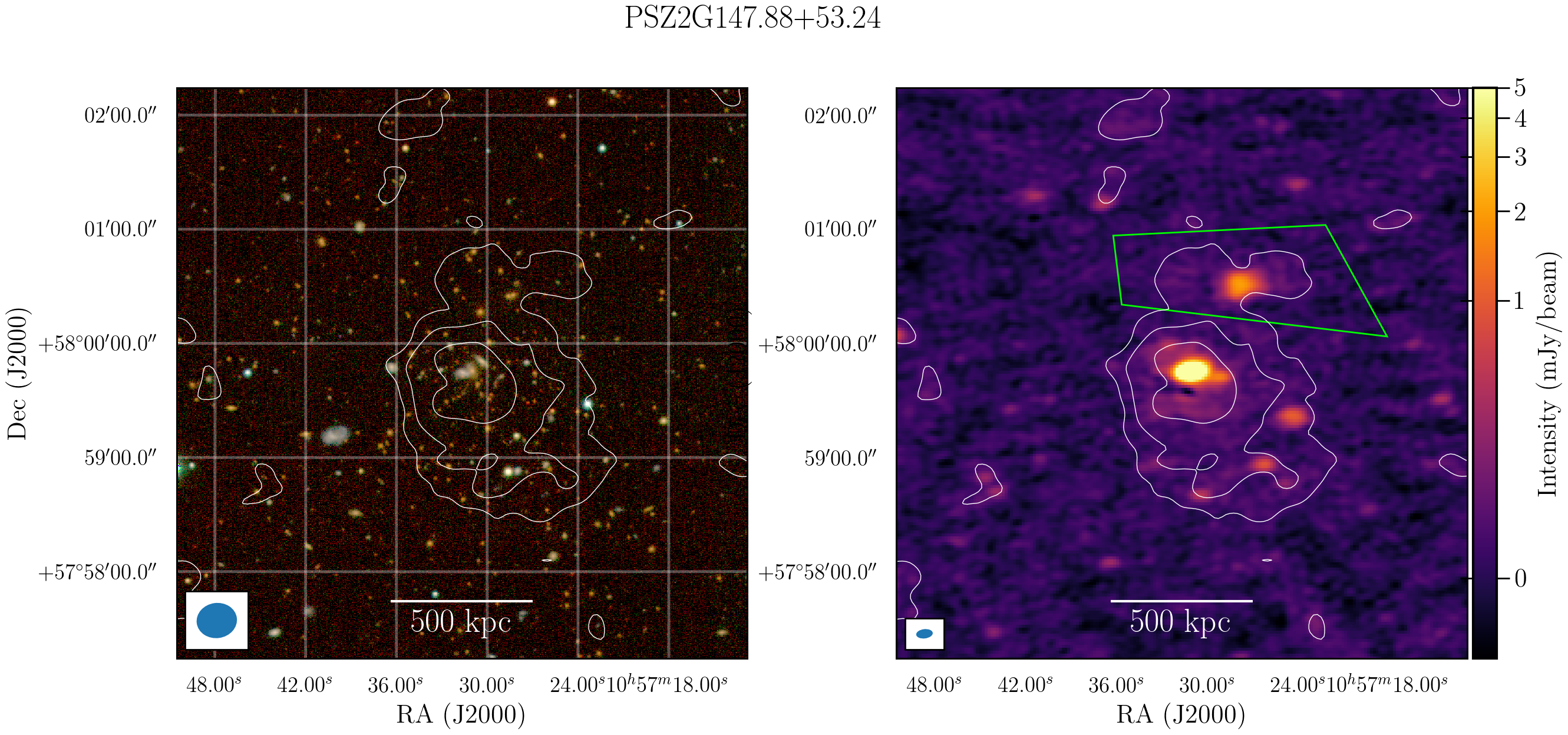}
    \caption{Low-resolution LOFAR diffuse emission from the cluster PSZ2G147.88+53.24 overlaid on $grz$ filters from the Legacy survey \citep{2019AJ....157..168D} (left) and overlaid on the high resolution radio intensity image (right). The contours are plotted at $[3,6,12]\sigma$, where $\sigma=83 \mu$\Jyb. The restoring beam sizes are $21''\times19''$ and $9''\times5''$ for the low-resolution and high-resolution radio images respectively. The green region marks the region masked from the fitting procedure.}
    \label{fig:PSZ2G147_opticalradio}
\end{figure*}

PSZ2 G147.88+53.24 is a massive, high-redshift \citep[$z=0.6$;][]{2016A&A...594A..27P} galaxy cluster. Diffuse emission has been recently reported by \citet{DiGenarro2020}, where the emission was classified as a giant radio halo. \citet{DiGenarro2020} measured a total flux density of $14.4 \pm 2.3$ mJy at 144 MHz by arithmetically subtracting the radio galaxies flux densities from the total flux density at low-resolution. 
The diffuse emission has a largest linear size of around 700 kpc. We employ a $uv$-cut corresponding to $400$ kpc at the cluster redshift (3447$\lambda$) for the compact source subtraction process. We confirm the detection of \citet{DiGenarro2020} in our deeper image, which is shown in Figure \ref{fig:PSZ2G147_opticalradio} where we show the low-resolution compact source subtracted radio contours overlaid on the $g,r,z$ optical image from the Legacy survey \citep{2019AJ....157..168D}. The high-resolution radio emission is shown the right panel of Figure \ref{fig:PSZ2G147_opticalradio}. The source to the north-west of the central radio galaxy in Figure \ref{fig:PSZ2G147_opticalradio} might be contributing to the low-resolution (compact source subtracted) radio contours, given the peculiar feature that is present in the low-resolution contours. Therefore, we decide to fit the halo with and without a mask covering the north-western source. The masked region is shown as the green region in Figure \ref{fig:PSZ2G147_opticalradio}.

As this cluster is at a high redshift, 50 kpc corresponds almost to the high-resolution beam size (at $z=0.6$, 50 kpc corresponds to 7.5''). Therefore, we taper to lower resolution, using a 10'' Gaussian taper, to make the fitting procedure converge better. The full width at half maximum of the restoring beam of the low-resolution image is $21.3''\times18.5^{\prime\prime}$. Without any masking, the best-fit parameters are $I_0=4.4\pm0.3$ $\mu$Jy arcsec$^{-2}$, $r_e$=$194\pm10$ kpc. Integrating the model in Equation \eqref{eq:circ} up to $2.6r_e$ with the best-fit parameters leads to a flux density at 144 MHz of $16.9 \pm 2.0$ mJy.
This corresponds to a 1.4 GHz power of $P_{1.4\mathrm{GHz}} = (1.1 \pm 0.4) \times 10^{24} $W Hz$^{-1}$ assuming $\alpha=-1.5\pm0.2$. 
When employing the mask shown in the right panel of Figure \ref{fig:PSZ2G147_opticalradio}, we find best-fit parameters $I_0=4.5\pm0.3$ $\mu$Jy arcsec$^{-2}$ and $r_e=186 \pm 11$ kpc, which correspond to a consistent integrated flux density of $16.0 \pm 2.0$ mJy at 144 MHz.

The resulting flux densities are a bit higher, but consistent within the error bounds with the value of $14.4 \pm 2.3$ mJy reported by \citet{DiGenarro2020}. This is to be expected, because our observations are deeper and Section \ref{sec:inject} showed that we are likely biased a bit high on the flux density values due to the compact source subtraction process. Manually measuring the flux within 3$\sigma$ contours results in a bit better agreement with a flux density of $14.7 \pm 1.6$ mJy.

\subsection{PSZ2G149.22+54.18}
\begin{figure}
    \centering
    \includegraphics[width=1.0\columnwidth]{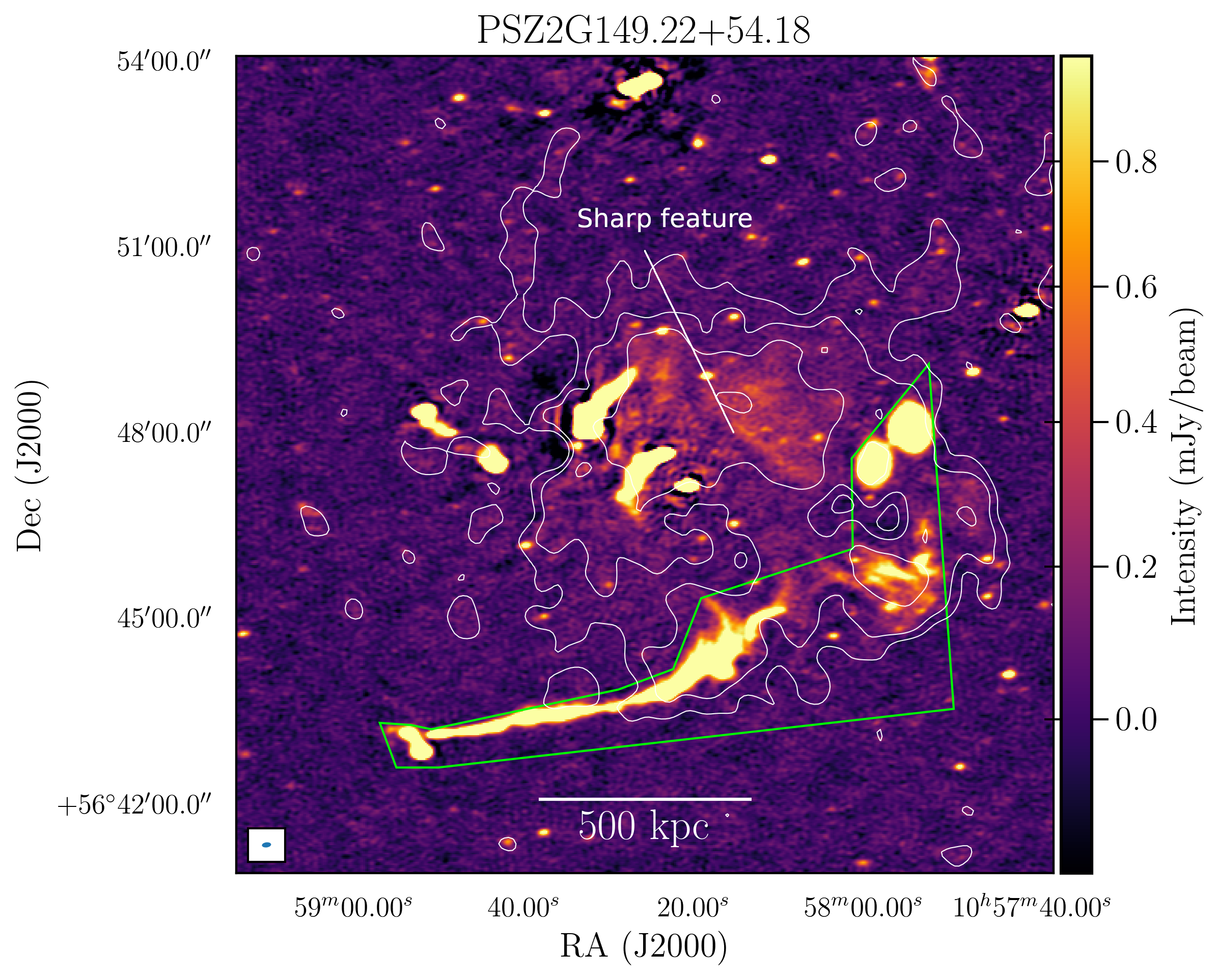}
    \caption{High-resolution ($9''\times5''$) LOFAR radio intensity image of Abell 1132 with low-resolution compact source subtracted contours at $[3,6,12,..]\sigma$, where $\sigma=149 \mu$\Jyb. The beam size of the low-resolution contours is $28''\times24''$. The green box indicates the region that is masked from the fitting procedure.}
    \label{fig:radioradioA1132}
\end{figure}

PSZ2G149.22+54.18, or Abell 1132 is a quite massive cluster, with a mass of 5.87$^{+0.23}_{-0.22}\times10^{14}M_\odot$ \citep{2016A&A...594A..27P} that is undergoing a merging event \citep{2015A&A...580A..97C}. It is located at a redshift of $z=0.1369$ \citep{1991ApJS...77..363S}. Diffuse emission was not picked up by previous VLA observations at 1.4 GHz \citep{2000NewA....5..335G}, but was clearly detected by previous observations with LOFAR \citep{Wilber_2017}. The central diffuse emission was classified as an ultra steep spectrum radio halo with $\alpha=-1.75 \pm 0.19$ between 144 and 325 MHz. The connection between the diffuse emission in the halo and the diffuse emission at the edge of the giant tailed radio galaxy was tentatively raised by \citet{Wilber_2017}, and is now clearly observed in the low-resolution contours shown in Figure \ref{fig:radioradioA1132}. We note that the halo size seems larger than previously determined, with the size inside the $3\sigma$ contours being $\sim 1.0$ Mpc $\times0.9$ Mpc in the east-west and north-south direction, respectively. 

To allow for a better comparison to the previous LOFAR observations, the compact source subtraction was done by using a $uv$-cut corresponding to 500 kpc at the cluster redshift (i.e., 1000$\lambda$ in the $uv$ plane). Since the giant head-tail radio galaxy blends in with the emission of the halo, we manually mask the tail from the fitting procedure. The mask is shown in the green box in Figure \ref{fig:radioradioA1132}.

The best-fit parameters are $5.7\pm0.1$ $\mu$Jy arcsec$^{-2}$ and $r_e = 235\pm4$ kpc. 
These correspond to a total flux density of the halo of $244.9 \pm 29.7$ mJy at 144 MHz, translating to $P_{1.4\mathrm{GHz}} = (2.5 \pm 0.3) \times10^{23}$W Hz$^{-1}$, assuming $\alpha=-1.75$, which is in agreement with the value reported by \citet{Wilber_2017}. 
The extent of the halo within the $3\sigma$ contour level is larger in our deep image than in the image of \citet{Wilber_2017}, which again points out that fitting the halo provides more robust flux density measurements than measuring the flux density within certain isophotes. 
Manually measuring the flux in the 3$\sigma$ contours (without a mask) results in a flux density of $261\pm31$ mJy, which is consistent with the flux from integrating the best-fit radial profile up to $2.6r_e$.
Although the halo is a bit more elongated in the east-west direction than in the north-south direction, the comparison with the manually measured flux density within 3$\sigma$ contours indicates that the circular model is still a reasonable assumption.

\subsection{PSZ2G084.69+42.28}

\begin{figure*}[tbh]
    \centering
    \includegraphics[width=1.0\textwidth]{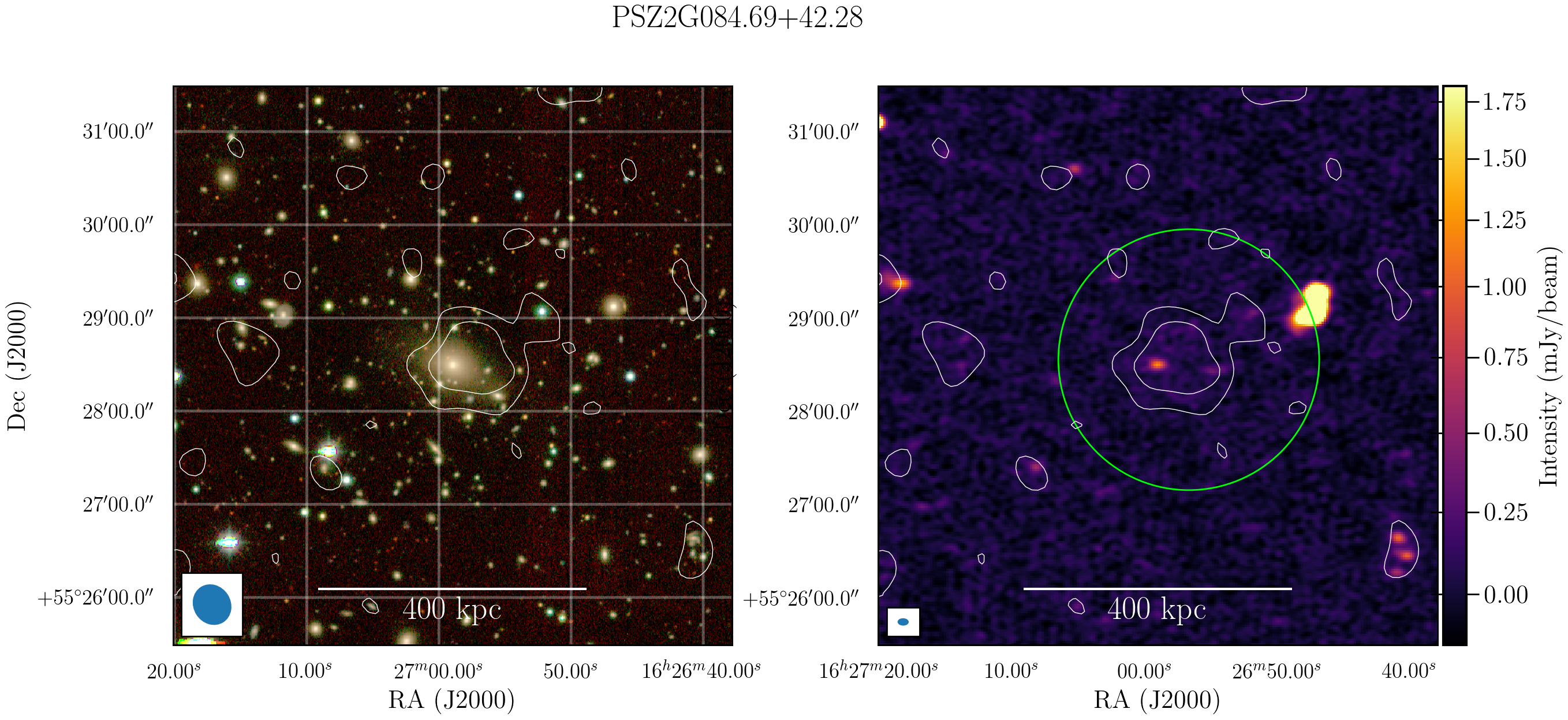}
    \caption{Low-resolution LOFAR diffuse emission from the cluster PSZ2G084.69+42.28 overlaid on $grz$ filters from the Legacy Survey (left) and overlaid on the high resolution radio intensity image (right). The contours are at $[2,4]\sigma$, where $\sigma = 152\mu$\Jyb. The high- and low-resolution beam sizes are $7''\times5''$ and $27''\times24''$ respectively. The fitting procedure was performed only within the region of the image contained by the green circle.}
    \label{fig:PSZ084both}
\end{figure*}

PSZ2G084.69+42.28 or Abell 2201, is a relatively low mass ($2.67_{-0.26}^{+0.27} \times10^{14} M_\odot$;\citealt{2016A&A...594A..27P}) galaxy cluster at a redshift of 0.13 \citep{2002ApJ...570..100L}, which has not been studied extensively. This cluster has the lowest mass estimate in our sample of PSZ clusters with a detection. We pick up weak diffuse emission from the cluster centre. This emission is visible in the low-resolution compact source subtracted contours overlaid on the high-resolution image in the left panel of Figure \ref{fig:PSZ084both}. The optical image overlay is shown in the right panel, which shows that the diffuse emission surrounds the brightest cluster galaxy (BCG).

Since the diffuse emission is very small in size, we employ a $uv$ cut of $2390\lambda$, corresponding to 200 kpc at the cluster redshift. Because the emission is relatively small in size, it is possible that it is AGN-related.
We enforce that the spherical profile is only fit in a region of approximately $400$ kpc by masking out the outer regions, because the emission is only barely picked up above the noise. This mask is shown in the right panel of Fig. \ref{fig:PSZ084both}.
The best-fit values are found to be $I_0=2.0^{+0.7}_{-0.6}$ $\mu$Jy arcsec$^{-2}$ and $r_e=57^{+18}_{-13}$ kpc. Integrating the analytical model up to $2.6r_e$ results in a total flux density of 5.5 $\pm$ 1.6 mJy at 144 MHz or a radio luminosity at 144 MHz of $(2.6\pm 0.8) \times 10^{23}$ W Hz$^{-1}$. 
Assuming a spectral index of $\alpha=-1.5 \pm 0.2$, we obtain a radio power of $P_{1.4\mathrm{GHz}}= (8.6 \pm 4.5) \times 10^{21}$W Hz$^{-1}$, 

Assuming $r_H/r_e=2.6$ \citep{10.1093/mnras/stx1475}, the radius of the diffuse emission is about 150 kpc, which is much smaller than typical radio halos and would imply a ratio $R_H/R_{500}=0.16$ that falls in the typical range of mini-halos \citep{2017ApJ...841...71G}. Thus based on the size we would identify this as AGN-related emission or a mini-halo. However, according to \citet{2007MNRAS.378.1565C} radio halos do not follow a self-similar scaling, with their size decreasing more rapidly than that of the hosting cluster with decreasing mass \citep[see also][]{radial_profile}. Thus, it is not unexpected that a radio halo would be smaller than halos found in high mass systems.

\subsection{MCXCJ1036.1+5713}\label{sec:MCXCJ1036}

\begin{figure*}[tbh]
    \centering
    \includegraphics[width=1.0\textwidth]{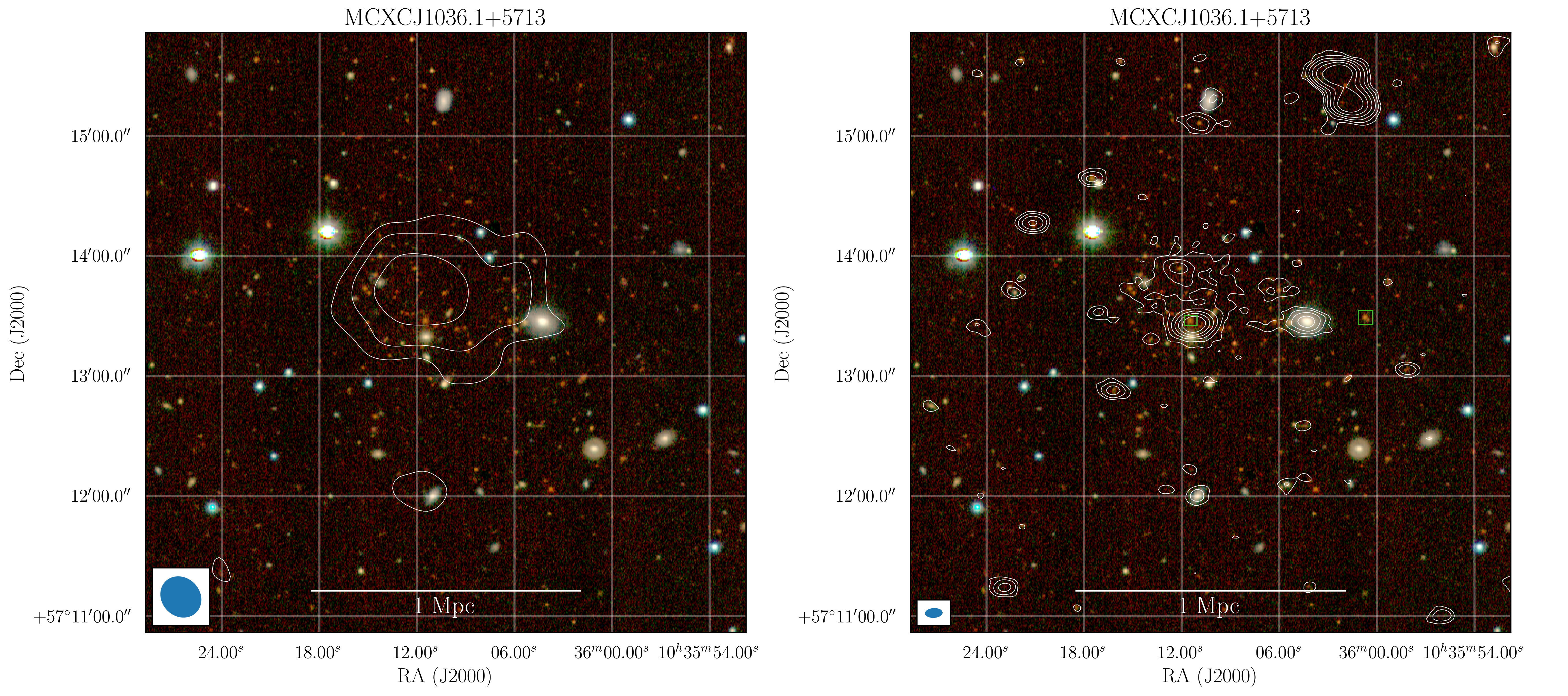}
    \caption{Optical image ($grz$ filters) of the cluster MCXCJ1036.1+5713 from the Legacy Survey with compact source subtracted low-resolution ($22''\times19''$) LOFAR contours overlaid (left) and high resolution compact source contours overlaid (right). Contours at $[3,6,12]\sigma$, where $\sigma$=86 $\mu$Jy beam$^{-1}$ and 38 $\mu$Jy beam$^{-1}$ respectively. The green boxes denote the galaxies with SDSS spectra available.}
    \label{fig:MCXCJ036_optical}
\end{figure*}

\begin{figure}
    \centering
    \includegraphics[width=1.0\columnwidth]{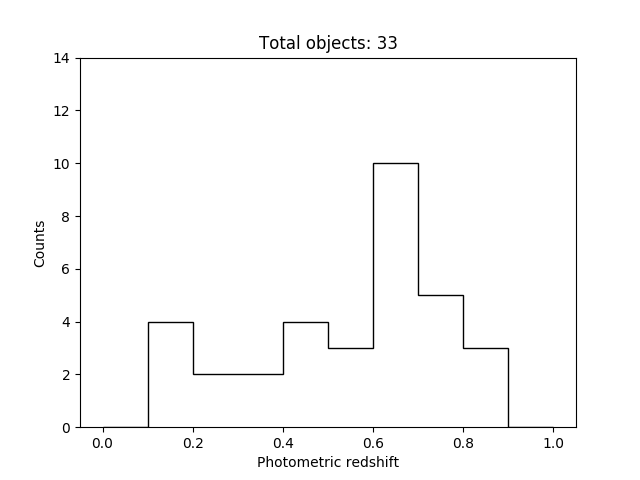}
    \caption{SDSS photometric redshifts of galaxies within a radius of roughly $1.5\prime$ from the central radio source of the cluster MCXCJ1036.1+5713.}
    \label{fig:histSDSS}
\end{figure}

This cluster was detected by the 400 deg$^2$ \textit{ROSAT} PSPC Galaxy Cluster Survey \citep{2007ApJS..172..561B} and lies at a redshift of $z=0.203$ according to \citet{2011A&A...534A.109P}. However, the optical image shown in Figure \ref{fig:MCXCJ036_optical} does not show a clear overdensity of low-redshift galaxies, but rather shows an overdensity of small, red galaxies, which suggests that the detected cluster lies at higher redshift. Figure \ref{fig:histSDSS} shows the SDSS photometric redshift estimates \citep{2020ApJS..249....3A} of galaxies within a radius of roughly $1.5\prime$ from the central radio source, and indeed an overdensity is apparent at $z=0.6-0.7$ rather than at $z=0.2-0.3$.
The optical counterpart to the central bright radio source shown in the high resolution contours (denoted by the green box) has a spectroscopic redshift of $z=0.76991$  \citep{2020ApJS..249....3A}. Another nearby source, which looks to be in the same cluster, also has a spectroscopic redshift of $z=0.76391$. 
For these reasons, we adopt a redshift of $z=0.76991$ for this cluster.

Correcting the X-ray luminosity given by the MCXC for this change in redshift, we find a mass estimate of $M_{500} = 3.3_{-1.7}^{+1.1} \times 10^{14} M_\odot.$ using the relation between $L_{500}$ and $M_{500}$ found by \citet{2010A&A...517A..92A}. The one sigma error reported is underestimated, as this only takes into account the intrinsic scatter in the $L_X-M$ relation.

While masses derived from X-ray luminosity are generally less well constrained than masses derived from the Sunyaev-Zel'dovich effect, the fact that the cluster is not present in the \textit{Planck} Sunyaev Zel'dovich catalogue \citep{2016A&A...594A..27P} can also be used to constrain the mass. From visual inspection of the Compton parameter maps released by \citet{2016A&A...594A..22P}, we note that there are various detections in a region of four degrees around this cluster, which makes it likely that the non-detection of this cluster is simply due to a low signal-to-noise ratio and thus low mass of the cluster.
The completeness of the PSZ2 catalogue as a function of mass and redshift \citep[Fig. 26 in][]{2016A&A...594A..27P} indicates that for the cluster redshift of $z=0.76991$, the catalogue is 50\% and 80\% complete for masses of $\sim 6.0$ and $\sim7.5$ $\times10^{14} M_\odot$ respectively. This provides us a fiducial upper limit to the mass of the cluster of $\sim 7.5\times10^{14} M_\odot$).


The compact source subtraction process for this cluster is shown in Figure \ref{fig:MCXCJ036}. The final panel shows a clear detection of extended diffuse radio emission, which can be best observed from the radio-optical overlay given in Figure \ref{fig:MCXCJ036_optical}. Because of the large size of this emission ($>800$ kpc), we classify this source as a radio halo. 

We find best-fit parameters $I_0 = 7.7 \pm 0.5 \mu$Jy arcsec$^{-2}$ and $r_e = 124\pm7$ kpc. Integrating the analytical profile given in Eq. \ref{eq:circ} results in a flux density of 9.8 $\pm$ 1.1 mJy. 
Assuming a spectral index of $-1.5\pm0.2$, this translates to a radio luminosity of $P_{1.4\mathrm{GHz}} = (1.2 \pm 0.4) \times10^{24}$ W Hz$^{-1}$ at 1.4 GHz\footnote{For completeness we note that assuming a redshift of z=0.203 would give a mass of $M_{500} = 0.88\times10^{14}M_\odot$ and a radio luminosity of $P_{1.4\mathrm{GHz}} = (3.8 \pm 1.7) \times10^{24}$ W Hz$^{-1}$}.

\subsection{SpARCS1049+56}\label{sec:SpARCS}
\begin{figure}
    \centering
    \includegraphics[width=1.0\columnwidth]{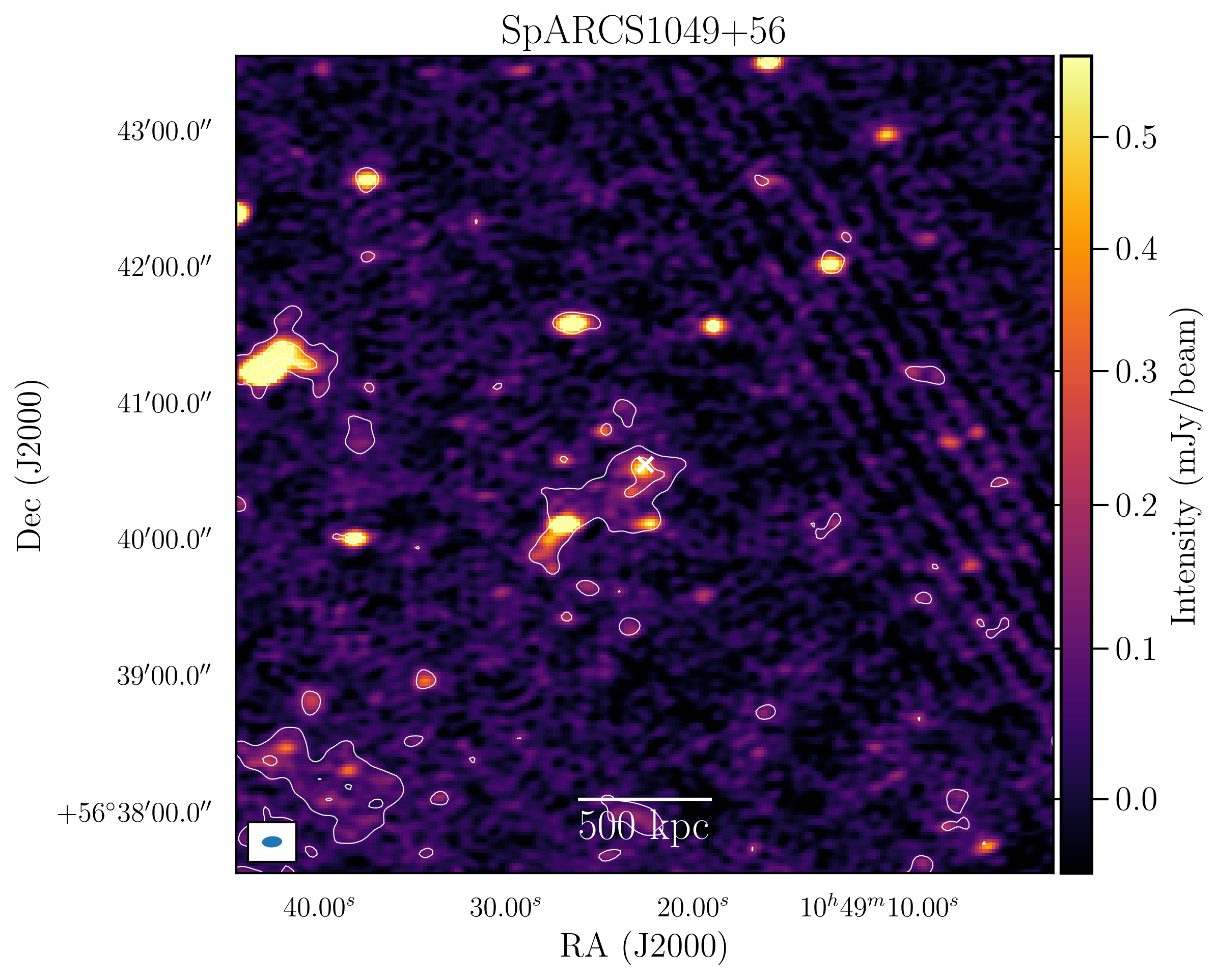}
    \caption{High-resolution ($9''\times5''$) LOFAR radio intensity image of SpARCS1049+56 with low-resolution ($14''\times9''$) compact source subtracted contours showing diffuse emission at $[3,6]\sigma$, where $\sigma=48 \mu$\Jyb. The white cross marks the location of the brightest cluster galaxy.}
    \label{fig:sparcsradioradio}
\end{figure}
SpARCS1049+56 is a very high redshift ($z=1.71$) cluster where star formation is actively taking place in the core, at a rate of $860 \pm 130$~ M$_\sun$~yr$^{-1}$ \citep{2015ApJ...809..173W}. The mass of the cluster was determined via infrared weak lensing to be $M_{200}= (3.5\pm1.2) \times10^{14}$~M$_\sun$ \citep{2020ApJ...893...10F}. 

We pick up some diffuse emission from this cluster. The compact source subtracted radio contours overlaid on the high-resolution radio map are shown in Fig. \ref{fig:sparcsradioradio}. The radio-optical overlay is shown in Appendix II, Figure \ref{fig:SPARCSoptical}. As the cluster is located at such a high redshift, it is difficult to properly subtract the compact sources from the diffuse component. It is clear that there is still some AGN-related emission contributing to the low-resolution contours, given the correlation between compact source locations and the location of the diffuse emission. 

We believe that the emission that is being picked up in the core is most likely AGN related, also because radio halos are expected to be intrinsically less luminous (by a factor of $B^2/B^2_{\mathrm{CMB}}$) with higher redshift due to inverse Compton losses \citep[e.g.,][]{2002A&A...396...83E,2006MNRAS.369.1577C,2019ApJ...881L..18C}. Even assuming a magnetic field of a few $\mu$G for the cluster at $z=1.71$ \citep[e.g.,][]{2019MNRAS.486..623D}, the synchrotron radiation would be reduced by about two orders of magnitude, making the detection of such a halo extremely unlikely by simple energetic arguments.

\subsection{SDSSC4-3094}
\begin{figure}[tbh]
    \centering
    \includegraphics[width=1.0\columnwidth]{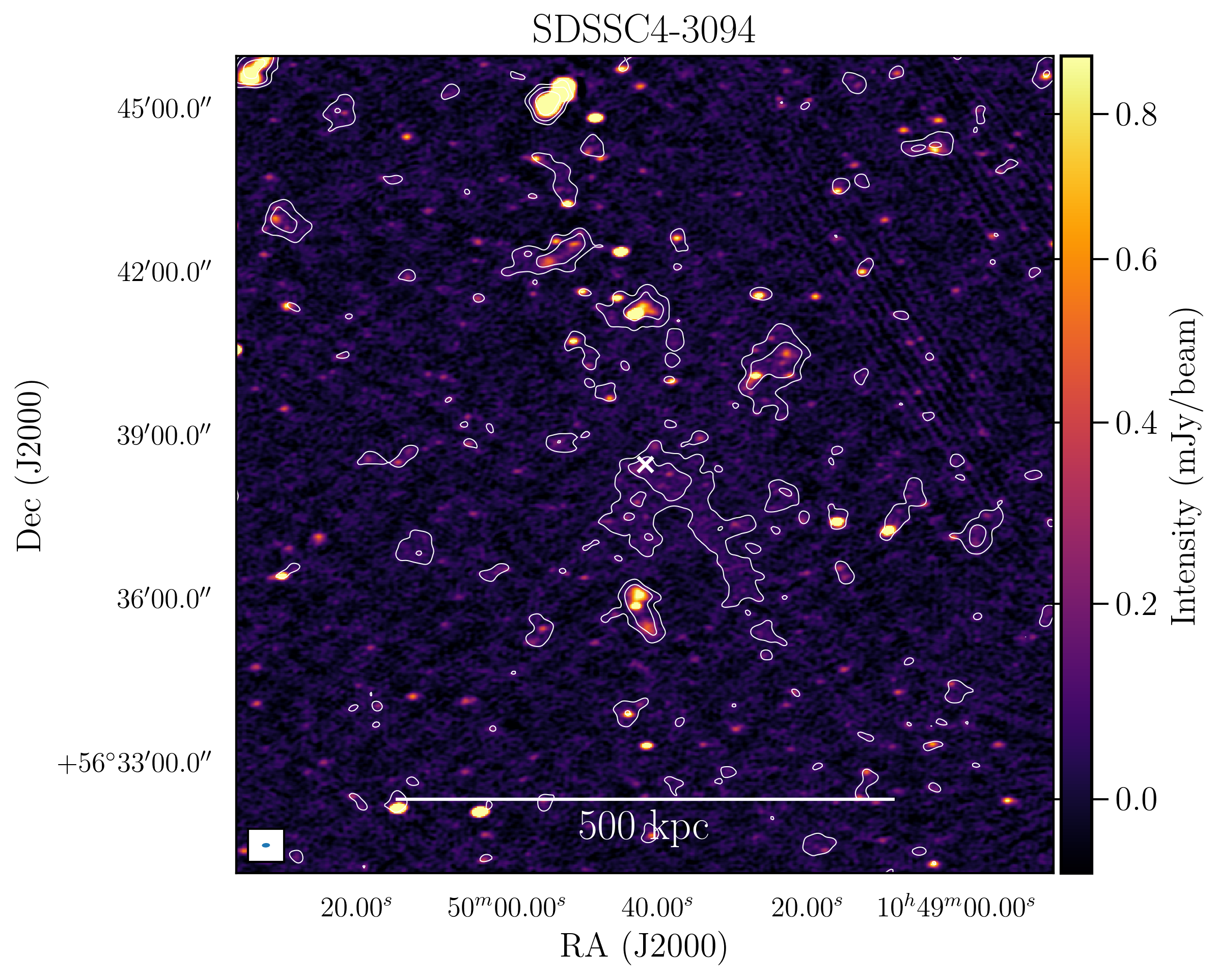}
    \caption{SDSSC4-3094 LOFAR radio intensity map with a restoring beam of $9''\times5''$, with low-resolution ($21''\times19''$) compact source subtracted contours showing diffuse emission at $[3,6,12..]\sigma$, where $\sigma=74 \mu$\Jyb. The white cross marks the location of the brightest cluster galaxy.}
    \label{fig:SDSS}
\end{figure}

A nearby galaxy cluster, SDSSC4-3094, identified in the Sloan Digital Sky Survey \citep{2005AJ....130..968M} at $z=0.04632\pm0.00083$ happens to be located in the same extracted region as SpARCS1049+56. The radio-optical overlay is shown in Appendix II (Figure \ref{fig:SDSSoptical}). From this cluster we detect diffuse emission shown in Fig. \ref{fig:SDSS} to the southwest of the BCG. This emission is not following the radio galaxy distribution and seems like genuine diffuse emission. However, it is likely not a radio halo, given the one-sided morphology. to the south-east. We classify this emission as remnant AGN emission due to this morphology and low surface brightness.

\subsection{Upper limits on non-detections}\label{sec:WHL}
The three WHL clusters that we identified showing possible diffuse emission are fairly unknown clusters. All radio-optical overlays for the WHL clusters are shown in Appendix II. Since these are optically detected clusters, we can estimate their mass from the richness. We use the relation given by \citet{2012ApJS..199...34W}
\begin{equation}
    \log M_{200} = (-1.49\pm0.05) + (1.17 \pm 0.03) \log R_{L^*}
\end{equation}
where $R_{L^*}$ is the cluster richness as reported in \citet{2012ApJS..199...34W} and $M_{200}$ is the mass in units of $10^{14}M_\odot$. To convert the masses to $M_{500}$ we use $M_{500}$ = 0.72 $M_{200}$, which assumes a Navarro–Frenk–White profile with a concentration parameter $c=5$ for the cluster scale dark matter halo \citep{1996ApJ...462..563N,2003MNRAS.342..163P}. The results of the subtract and extract procedure are briefly stated per cluster.
\begin{figure}
    \centering
    \includegraphics[width=1.0\columnwidth]{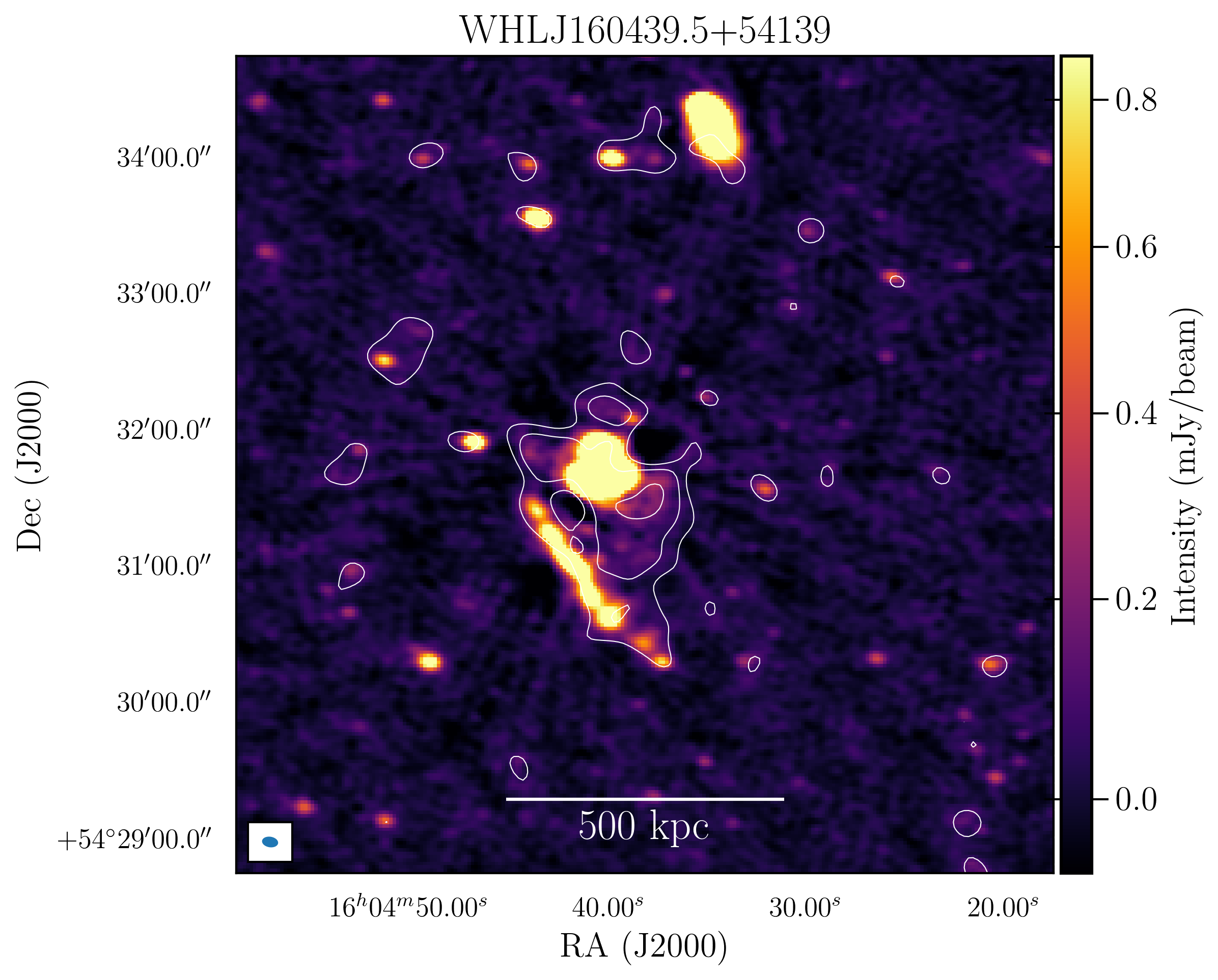}
    \caption{WHLJ160439.5+543139 high-resolution ($7''\times5''$) LOFAR radio intensity image with low-resolution ($17''\times15''$) compact source subtracted contours at $[3,6,12,..]\sigma$, where $\sigma=72 \mu$\Jyb.}
    \label{fig:1604radrad}
\end{figure}

\begin{figure*}
    \centering
    \includegraphics[width=0.8\textwidth]{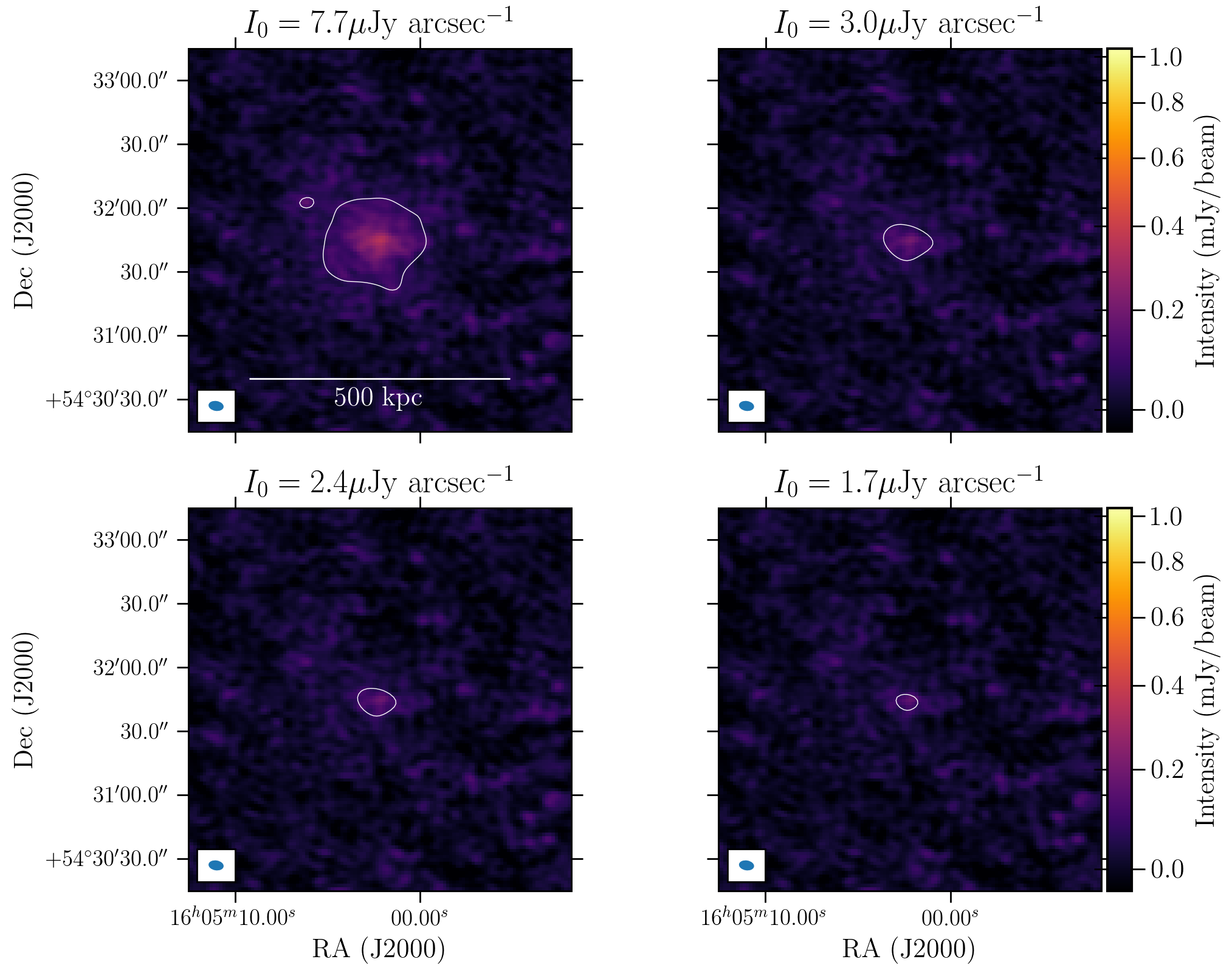}
    \caption{The injected mock halos with decreasing surface brightness near the cluster WHLJ160439.5+543139. The contours are plotted at 3$\sigma$, where $\sigma=25 \mu$\Jyb. The halo is still defined as detected in the bottom left panel, but not defined as detected in the bottom right panel.}
    \label{fig:uplimWHL1}
\end{figure*}

\textbf{WHLJ160439.5+543139}
This cluster is located at a redshift of $z=0.2655$ and shows a head-tail radio galaxy to the south-east of the likely BCG (Fig. \ref{fig:1604radrad}). From the richness we estimate a mass $M_{500} = (2.95 \pm 0.50 )\times 10^{14} M_\odot$. We tentatively detect diffuse emission surrounding the BCG, but due to the complexity of the emission from a head-tail radio galaxy, the AGN emission cannot be fully subtracted. To provide an upper limit on a halo detection, we inject halos slightly east of the cluster, in a region without contaminating radio sources. Following the correlations mentioned in Section \ref{sec:MeasProperties}, we initially set $I_0=7.7\mu$\Jya and $r_e=65$ kpc. We find that the halo is easily detected for these values, as shown in the first panel of Figure \ref{fig:uplimWHL1}. The resulting upper limit is found for $I_0=2.4\mu$\Jya, as is shown in the bottom left panel of Figure \ref{fig:uplimWHL1}. This results in an integrated flux density of $3.0$ mJy, which translates to an upper limit on the radio power at 144 MHz of $P_{\mathrm{144MHz}}=7.4\times10^{23}$ W Hz$^{-1}$.

\textbf{WHLJ161135.9+541635}
\begin{figure}
    \centering
    \includegraphics[width=1.0\columnwidth]{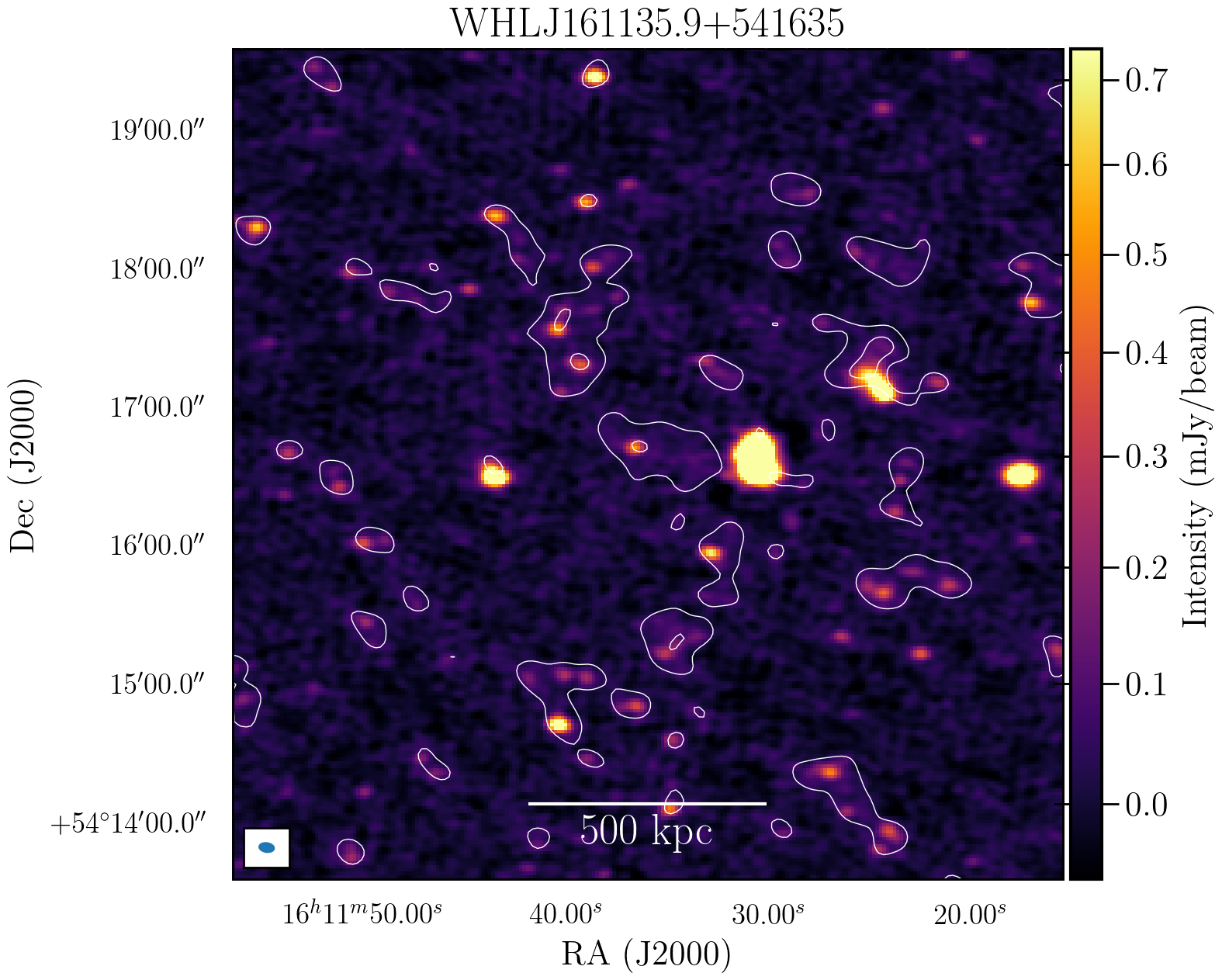}
    \caption{WHLJ161135.9+541635 high-resolution ($7''\times5''$) LOFAR radio intensity image with low-resolution ($20''\times18''$) compact source subtracted contours at $[3,6,12,..]\sigma$, where $\sigma=62 \mu$\Jyb}
    \label{fig:1135radrad}
\end{figure}
This cluster lies at a redshift of $z=0.3407$, with an estimated mass of $M_{500} = (3.4 \pm 0.6 )\times 10^{14} M_\odot$ and was selected visually because there seemed to be diffuse emission around the likely BCG in the wide-field image. However, after extraction and subtraction, no diffuse emission was detected, as shown in Figure \ref{fig:1135radrad}. For this cluster we find an upper limit for the values $I_0=3.5\mu$\Jya and $r_e=80$ kpc, corresponding to a total flux density of $4.4$ mJy. This translates to an upper limit of $P_{\mathrm{144MHz}}=2.0\times10^{24}$ W Hz$^{-1}$.

\textbf{WHLJ161420.1+544254}
\begin{figure}
    \centering
    \includegraphics[width=1.0\columnwidth]{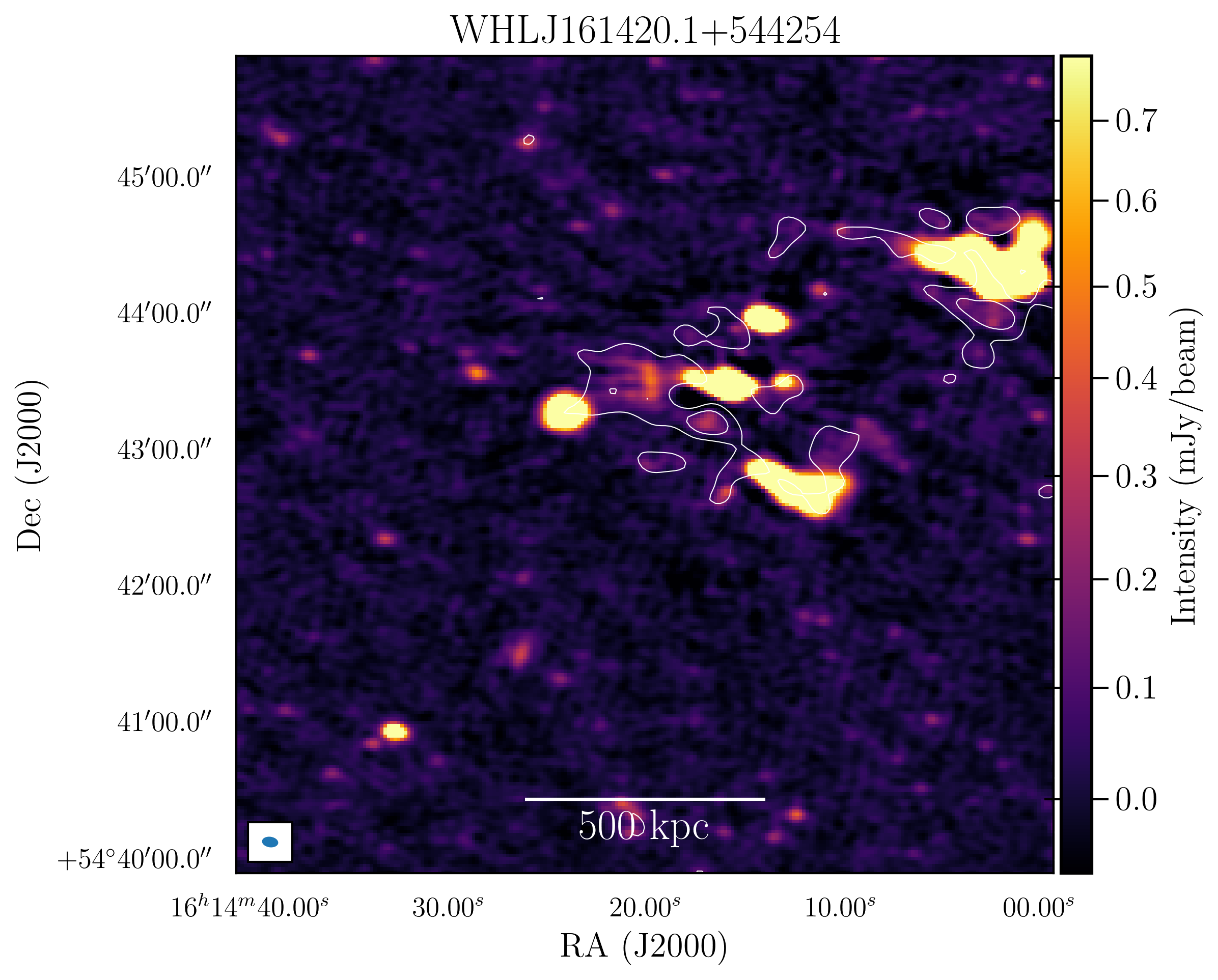}
    \caption{WHLJ161420.1+544254 high-resolution ($7''\times5''$) LOFAR radio intensity image with low-resolution ($16''\times13''$) compact source subtracted contours at $[3,6,12,..]\sigma$, where $\sigma=66 \mu$\Jyb}
    \label{fig:161420radrad}
\end{figure}
This WHL cluster also has a rather complex radio morphology, which combined with the leftover calibration artefacts prohibited the clear separation of AGN and diffuse emission (Fig. \ref{fig:161420radrad}). The cluster lies at a redshift of $z=0.3273$ with a mass of $M_{500} = (2.85 \pm 0.50 )\times 10^{14} M_\odot$. For this cluster we find an upper limit for the values $I_0=3.1\mu$\Jya and $r_e=64$ kpc, corresponding to an integrated flux density of 2.6 mJy and a radio power of $P_{\mathrm{144MHz}}=1.1\times10^{24}$ W Hz$^{-1}$.

\begin{figure}
    \centering
    \includegraphics[width=1.0\columnwidth]{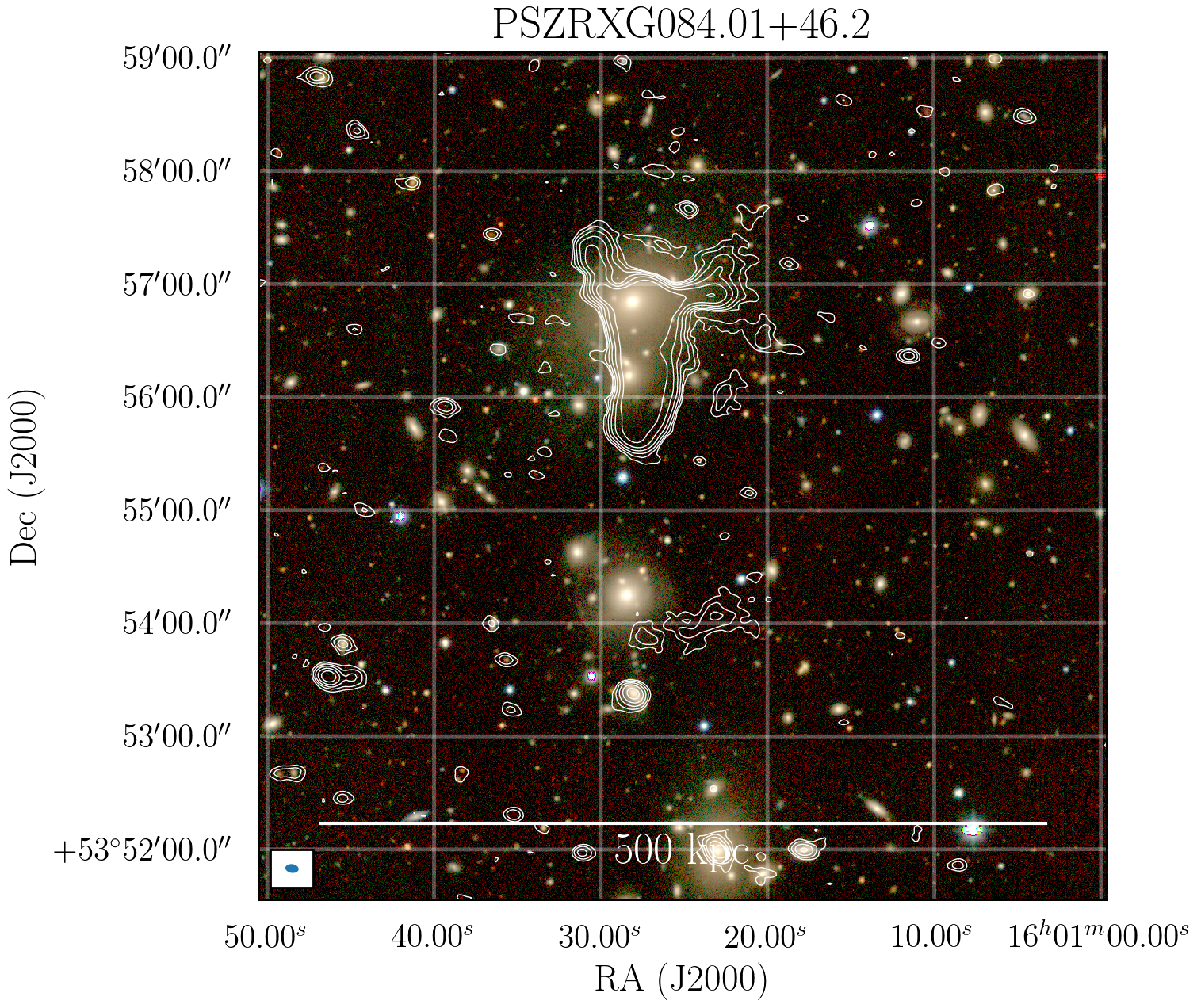}
    \caption{Optical ($grz$ filters) image from the Legacy survey with high-resolution ($7''\times5''$) LOFAR radio contours overlaid for the cluster PSZRXG084.01+46.28. Contours are spaced at $[3,6,12,..]\sigma$ where $\sigma=39\mu$\Jyb.}
    \label{fig:PSZRXopt}
\end{figure}

\textbf{PSZRX G084.01+46.28} or Abell 2149 is quoted to have a redshift of 0.1068 in the PSZ2 and MCXC catalogues, however it has been identified as a duplicate cluster with a redshift measurement discrepancy of more than 10 per cent in the MCXC catalogue \citep[see table B.1. of][]{2011A&A...534A.109P}. \citet{2006AJ....132.1275R} also noted the discrepancy between the redshift of 0.1068 quoted by the NORAS catalogue \citep{2000ApJS..129..435B} and $z=0.0675$ in the eBCS catalogue \citep{2000MNRAS.318..333E}. They noticed that the X-ray peak of the RASS image lies near an apparent BCG at the lower redshift. We adopt for this source the lower redshift of 0.0675 as well, since, as is shown in Figure \ref{fig:PSZRXopt} the radio emission is also concentrated around the brightest cluster galaxy (BCG; at coordinates 16h01m28.10s +53$^\circ$56m:50.8s) which is located at a redshift of $z=0.06544$ \citep{2013yCat.5139....0A}. 

Since the redshift was overestimated, the mass of this cluster is overestimated as well. We calculate the corrected mass by assuming $z=0.0675$ and interpolating the mass-redshift degeneracy curve given by the ComPRASS catalogue \citep{2019A&A...626A...7T}. This results in a corrected cluster mass of $M_{500}=1.37^{+0.25}_{-0.27} \times 10^{14} M_\odot$.

This cluster is a difficult case since there is extended AGN emission surrounding the BCG, with a peculiar, bull head-like shape. Therefore, the central part of a radio halo would be obscured. However, the bull-head feature is quite narrow and we see no clear extended emission outside of it.
We derive an upper limit by injection of a mock halo close to the cluster. The value of $r_e$ found by following the correlations mentioned in Section \ref{sec:MeasProperties} is only $25$ kpc because of the low mass of this cluster. We choose to set $r_e=65$ kpc, as diffuse sources with an $r_e$ of 25 kpc would generally not be classified as a radio halo. Setting a larger $e$-folding radius results a more conservative upper limit. The upper limit for the peak surface brightness $I_0$ is found to be 2.5$\mu$\Jya, which results in a total flux density of 29 mJy or a radio power of $P_{\mathrm{144MHz}}=3.3\times10^{23}$ W Hz$^{-1}$.

\textbf{MCXC J1033.8+5703}
\begin{figure}
    \centering
    \includegraphics[width=1.0\columnwidth]{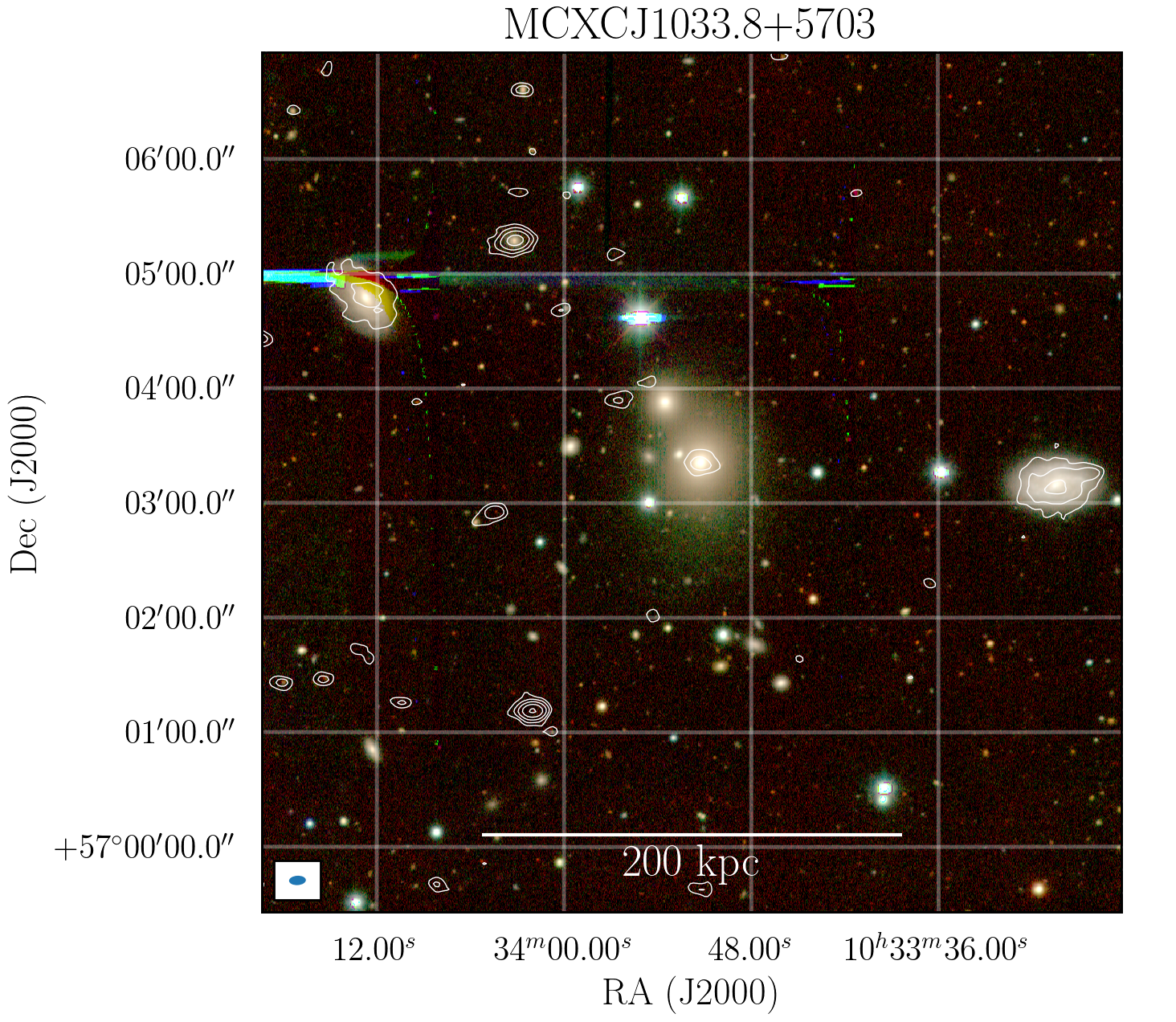}
    \caption{Optical ($grz$ filters) image from the Legacy survey with high-resolution ($9''\times5''$) LOFAR radio contours overlaid for the cluster MCXCJ1033.8+5703. Contours are spaced at $[3,6,12,..]\sigma$ where $\sigma=48\mu$\Jyb.}
    \label{fig:MCXCJ1033}
\end{figure}
No diffuse emission is picked up from this cluster (Fig. \ref{fig:MCXCJ1033} shows the optical emission with overlaid radio contours), which is not unexpected given the low mass of $M_{500}=0.128\times10^{14} M_\odot$ \citep{2011A&A...534A.109P}. If the mass is correct, this particular source is closer to a galaxy group than a galaxy cluster. Some galaxy groups have detected extended synchrotron emission, but their origin is not fully clear \citep[e.g.,][]{2011ApJ...732...95G,2017A&A...603A..97N,2019A&A...622A..23N}. Because is it unknown whether such low mass configurations of galaxies can host radio halos, we do not provide an upper limit for this cluster.

\textbf{MCXC J1053.3+5720}
\begin{figure}
    \centering
    \includegraphics[width=1.0\columnwidth]{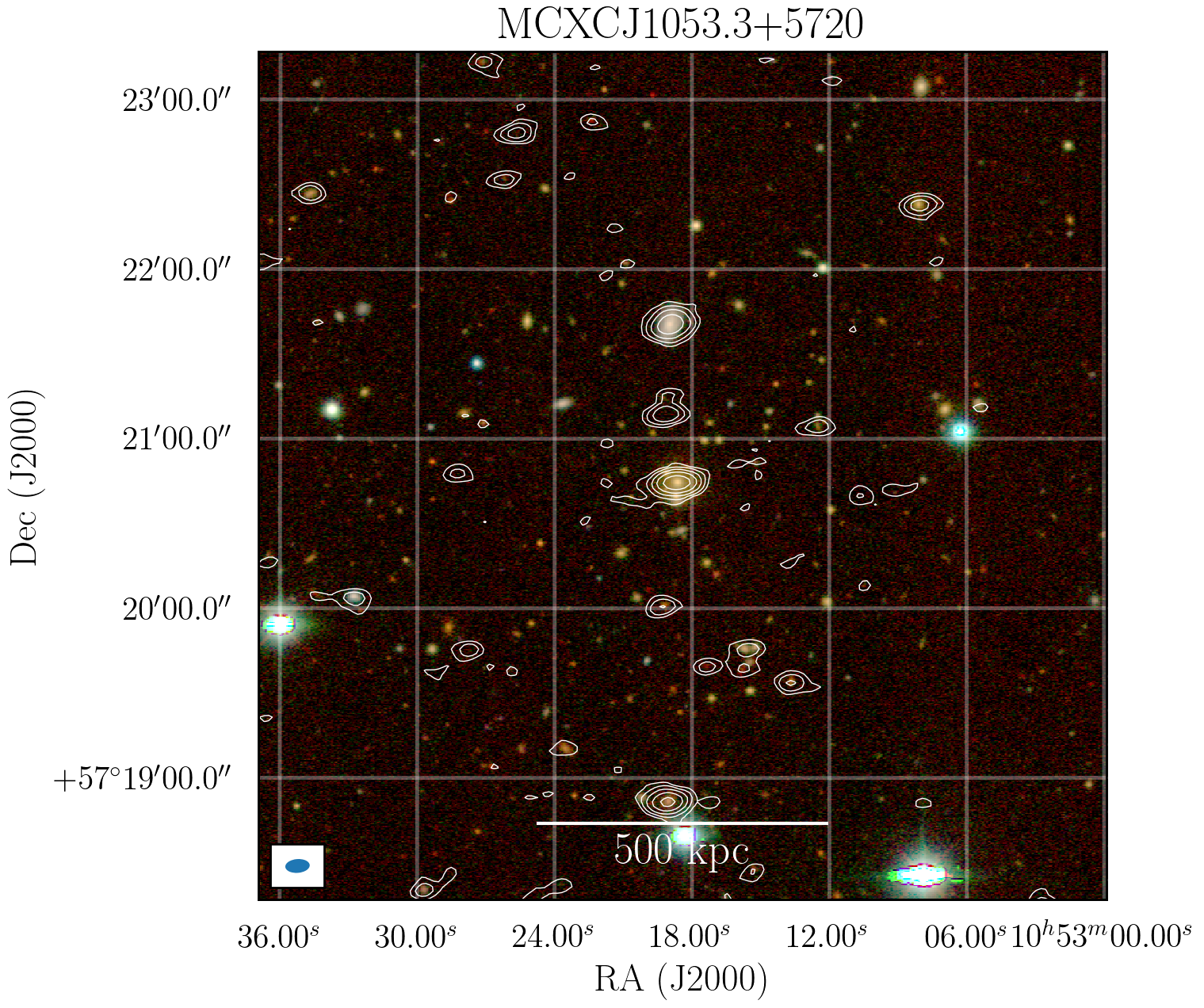}
    \caption{Optical ($grz$ filters) image from the Legacy survey with high-resolution ($9''\times5''$) LOFAR radio contours overlaid for the cluster MCXCJ1053.3+5720. Contours are spaced at $[3,6,12,..]\sigma$ where $\sigma=32\mu$\Jyb. }
    \label{fig:MCXCJ1053}
\end{figure}
This cluster is quite a low mass cluster according to the MCXC derived mass of $M_{500} = 0.487\times10^{14}M_\odot$ \citep{2011A&A...534A.109P}. It shows no diffuse emission, as is expected from such a low mass cluster. Although the mass is at least above a few times $10^{13}M_\odot$, the MCXC derived mass is still about a factor of five lower than the lowest mass cluster with a halo detection. Therefore, we do not consider it informative to provide an upper limit on such a cluster.

\begin{table*}[]
\centering
\caption{Resulting best-fit values for $I_0$ and $r_e$, as well as the total flux density $S_\nu^{\mathrm{fit}}$ from integrating the spherical model with the best-fit values. Parameter $d$ denotes half of the largest linear size of the $3\sigma$ contours in $e$-folding radii, and $S_{\nu}^{\mathrm{dfit}}$ is the flux density obtained by integrating up to that value of $d$. $S_{\nu}^{\mathrm{3}\sigma}$ is the flux density measured within 3$\sigma$ (2$\sigma$ for PSZ2G084.69+42.28) contours and $S_{\nu}^{\mathrm{2.6}r_e}$ is the flux density measured in a spherical region with a radius of $2.6r_e$.} 
\label{tab:comparisons}
\resizebox{\textwidth}{!}{%
\begin{tabular}{@{}lllllllll@{}}
\toprule
Source Name                 & $I_0${[}$\mu$Jy arcsec$^{-2}${]} & $r_e${[}kpc{]}    & $S_{\nu}^{\mathrm{fit}}$ & $d$   & $S_{\nu}^{\mathrm{dfit}}$ & $S_{\nu}^{\mathrm{3}\sigma}$ & $S_{\nu}^{\mathrm{2.6}r_e}$ & $\chi^2_{\mathrm{red}}$ \\ \midrule
PSZ2G147.88+53.24           & $4.4_{-0.3}^{+0.3}$              & $194_{-10}^{+11}$ & $16.9 \pm 2.0$           & $2.5$ & $16.4 \pm 2.0$            & $14.7 \pm 1.6$               & $16.0 \pm 1.8$              & 0.79                    \\
PSZ2G147.88+53.24 (mask)      & $4.5_{-0.3}^{+0.3}$              & $186_{-11}^{+11}$ & $16.0 \pm 2.0$           & $2.2$ & $14.0 \pm 1.7$            & $13.0 \pm 1.4$               & $15.5 \pm 1.7$              & 0.79                    \\
PSZ2G149.22+54.18           & $5.7_{-0.1}^{+0.1}$              & $235_{-3}^{+4}$   & $244.9 \pm 29.7$         & $2.5$ & $236.8 \pm 29.0$          & $261.0 \pm 30.6$             & $271.3 \pm 31.7$            & 1.24                    \\
PSZ2G084.69+42.28 & $2.0_{-0.6}^{+0.7}$              & $57_{-13}^{+18}$  & $5.5 \pm 1.6$            & $1.9$ & $4.3 \pm 1.2$             & $3.4 \pm 0.5$                & $4.7 \pm 0.8$               & 0.05                    \\
MCXCJ1036.1+5713            & $7.7_{-0.5}^{+0.5}$              & $124_{-6}^{+7}$   & $9.8 \pm 1.1$            & $3.6$ & $11.7 \pm 1.3$            & $9.9 \pm 1.1$                & $8.6 \pm 0.9$               & 0.57                    \\ \bottomrule
\end{tabular}
}
\end{table*}

\section{Discussion}\label{sec:discussion}
\begin{figure}
    \centering
    \includegraphics[width=1.0\columnwidth]{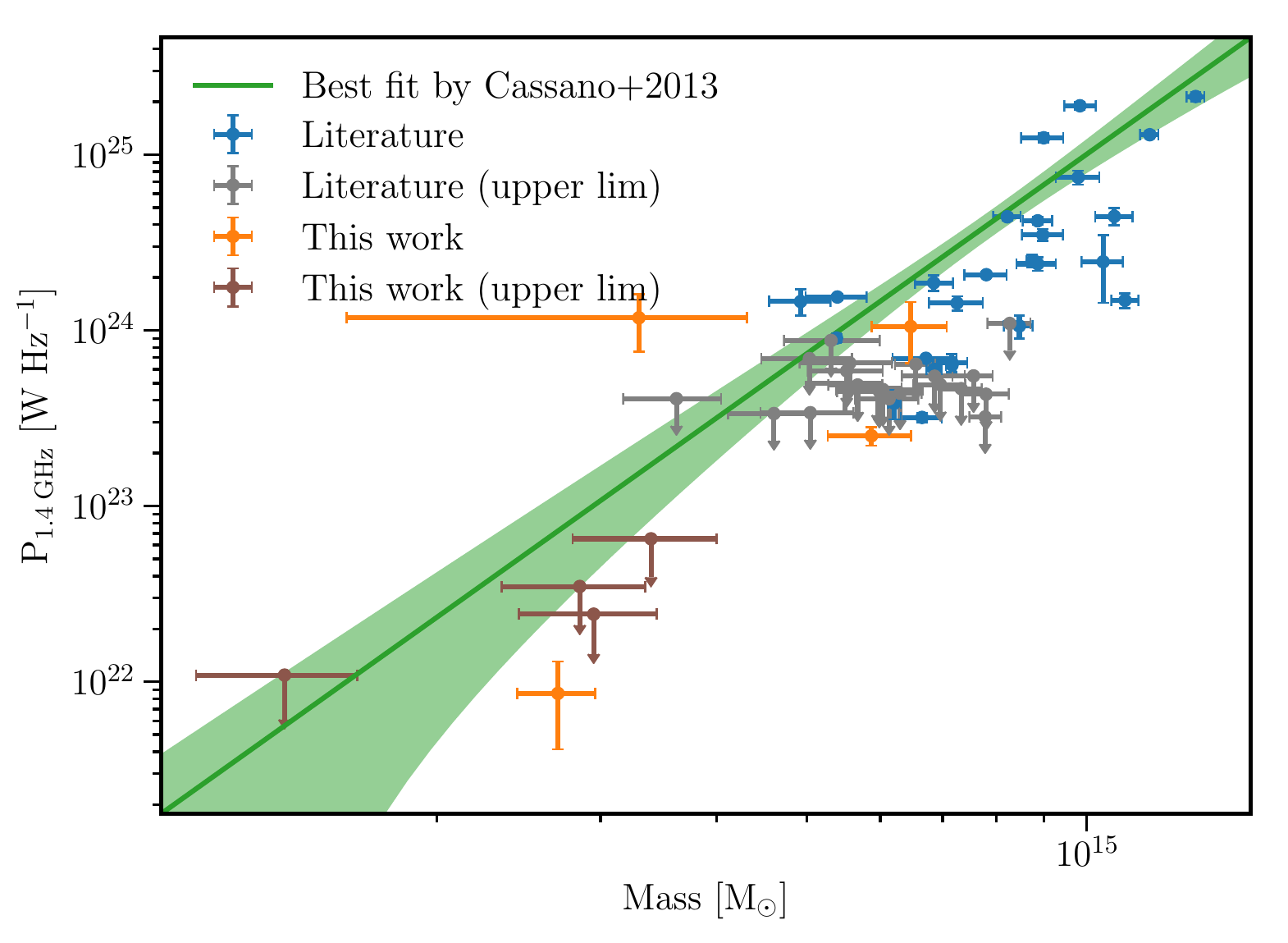}
    \caption{The radio halo power - mass diagram for the clusters in this work and a sample of clusters from \citet{WeerenReview} (Fig. 9) and references therein. The best-fit relation for radio halos, with the shaded 95\% confidence region, from \citet{Cassano2013} is shown in green. Note that individual halos can easily scatter outside of the shaded region due to the intrinsic scatter of host cluster properties. For the sources in this work where the spectral index is unknown, we have assumed $\alpha=-1.5\pm0.2$.}
    \label{fig:correlation}
\end{figure}
To quantify the robustness of the fitting procedure used to estimate the flux density of the diffuse emission in this paper, we compare the fitted flux densities to flux densities measured in various other ways. Table \ref{tab:comparisons} shows the best-fit values of $I_0$ and $r_e$, as well as a comparison with the flux density measured within 3$\sigma$ contours ($S_{\nu}^{\mathrm{3}\sigma}$) and the flux density measured in a spherical region with a radius of $2.6r_e$ ($S_{\nu}^{\mathrm{2.6}r_e}$). We find generally good agreement of the flux density measured within $3\sigma$ contours and the flux density from integrating the best-fit model, indicating that setting $d=2.6$ and using a circular model are reasonable choices for the clusters presented here. The only outlier is the source MCXCJ1036.1+5713, where the $3\sigma$ contours extend beyond $2.6e$-folding radii, and the actual flux density of the diffuse emission is thus slightly larger than the value of $9.8\pm1.1$ that is found by integration of the best-fit model. Integrating the model up to the value of $d=3.6$ results in a flux density of $11.7\pm1.3$ mJy.

We have found diffuse emission from three galaxy clusters in this study. One cluster hosts a new high-redshift radio halo, PSZ2G084.69+42.28, and two have been observed previously with shallower LOFAR observations, PSZ2 G147.88+53.24 and PSZ2 G149.22+54.18. We tentatively detect diffuse emission from the cluster PSZ2G084.69+42.28, but this has to be confirmed with upcoming deeper data releases. 
Upper limits have been put on the clusters PSZRX G084.01+46.28, WHL J160439.5+543139, WHL J161135.9+541635 and WHL J161420.1+544254. These results are compared to the well-known scaling relation between radio halo power and cluster mass from \citet{Cassano2013} derived for massive ($M_{500}>5\times10^{14}$) clusters. This is shown in Figure \ref{fig:correlation}. We find that the radio power of the diffuse emission in the low mass clusters PSZ2G084.69+42.28 and MCXCJ1036.1+5713 are inconsistent with the statistical error on the the best fit radio halo power - cluster mass correlation observed for higher mass systems. However, given the fact that radio halos are expected to scatter intrinsically around the correlation due to the different intrinsic properties of galaxy clusters and the different properties of mergers \citep{2009A&A...507..661B}, two data points are not yet enough to conclude a significant deviation. 

The turbulent re-acceleration model states that radio halos are caused by merger-induced turbulence in the intra-cluster medium (ICM) which re-accelerates relativistic electrons. A key prediction of this model is that lower mass clusters have less energetic merger events and thus less turbulent energy is being transferred to accelerate particles, leading to less powerful and steeper spectra radio halos \citep{BrunettiJones}. These halos can only be picked up by sensitive low-frequency instruments. Calculations based on the turbulent re-acceleration model predict 1000-3000 halos with an integrated flux density at 150 MHz of 10 mJy in the whole sky \citep{2006MNRAS.369.1577C,2019ApJ...879..104L}. The three fields considered in this work cover an area of about 60 deg$^2$, thus we would expect about 3 radio halo detections above an integrated flux density of 10 mJy. The results presented in this work are in line with these predictions. 

Our study shows the potential of deep LOFAR observations to detect diffuse emission from galaxy clusters with masses below $5\times10^{14}M_\odot$, thus entering a poorly explored territory. In the 8 hour LOFAR observations from the LOFAR Two Metre Sky Survey \citep{2019A&A...622A...1S}, the diffuse emission in PSZ2G084.69+42.28 is undetected and the diffuse emission in MCXCJ1036.1+5713 is barely detectable. Diffuse emission in a few other low mass clusters have been detected previously with LOFAR \citep[e.g.,][]{2016MNRAS.459..277S,2019A&A...622A..21H,2019A&A...630A..77B,2020A&A...634A...4M}. The diffuse emission found in PSZ2G084.69+42.28 and MCXCJ1036.1+5713 are important additions to the sparse sample of low mass $(<5\times10^{14}M_\odot$) clusters.

Theoretically, due to the lower turbulent energy budget in these low mass systems, the contribution from secondary electrons from hadronic collisions may become the dominant mechanism for powering radio halos \citep[e.g.,][]{2012A&A...548A.100C}. The transition from re-acceleration to hadronic halos depends on several unknowns, such as the energy budget of cosmic-ray protons (CRp) in clusters and the extension of the regions where turbulent energy is dissipated into re-acceleration of particles. Models that assume that the energy budget of CRp is at the levels constrained by \textit{Fermi-LAT} upper limits and that turbulence is dissipated in Mpc$^3$ regions (independent of cluster mass) predict a transition to hadronic halos at typical 150 MHz luminosities of $\sim10^{24}$W Hz$^{-1}$ \citep[e.g.,][]{2012A&A...548A.100C}. This value is similar to the radio luminosity found in PSZ2G084.69+42.28 and the upper limits obtained in this study, showing that deep observations with LOFAR can potentially constrain this transition.  

To investigate the possibility of a transition observationally, it is important to determine the dynamical state of the studied clusters. If radio halos in low mass clusters are still strongly connected to merger events, then that would suggest that the re-acceleration model still plays the dominant role, with implications on the extension of the turbulent regions and on the energy budget of CRp. The dynamical state of PSZ2G084.69+42.28 is also important to properly classify the diffuse emission.

Higher frequency follow-up observations are useful to differentiate between the two particle acceleration mechanisms. We have checked the ancillary 610 MHz GMRT observations of the ELAIS-N1 and Lockman Hole fields taken by  \citet{2008MNRAS.383...75G,2008MNRAS.387.1037G}, but unfortunately all PSZ2 and MCXC sources are just outside of the field-of-view of the GMRT observations. The WHL sources are observed, but show no sign of diffuse emission in the GMRT images. If, in future studies the spectral index of radio halos in low-mass sources is found to be very steep $\alpha \lessapprox -1.5$ a significant hadronic contribution will be ruled out \citep[e.g.,][]{2004A&A...413...17P,2008Natur.455..944B}.

The discovery of a radio halo in MCXCJ1036.1+5713 is particularly intriguing due to the combination of relatively low mass ($M_{500} \sim 3.3\times10^{14}$) and high redshift ($z=0.76991$). Models predict a gradual decline of the fraction of clusters with radio halos at high redshift \citep[e.g.,][]{2006MNRAS.369.1577C}. The observed decline is less prominent at low frequencies due to the increasing population of very steep spectrum halos that are expected to be more common at high redshift. Depending on the clusters magnetic field strength, a fraction of halos up to 10-25\% in clusters with $M_{500}\sim (3-4) \times 10^{14} M_\odot$ at a redshift of 0.7 is predicted to be observed with LOFAR \citep{2019ApJ...881L..18C}.  Better X-ray data with modern telescopes are needed to obtain a good estimate of the cluster mass and dynamical state.

Finally, our deeper images confirm that PSZ2G149.22+54.18 (Abell 1132) is hosting an under-luminous and steep-spectrum radio halo, which supports the idea that Abell 1132 is in a late merger state with weak turbulence \citep{Wilber_2017}. 
Due to the high sensitivity of the current data, we see the halo emission blending with the outer edge of the giant head-tail radio galaxy. The possibility has been raised that gently re-energized tails \citep[GreETs;][]{deGasperine1701634} can provide a seed population of relativistic electrons for the generation of the cluster-scale emission.
The interplay between the giant head-tail radio galaxy and radio halo seems to corroborate this scenario, although observations at different frequencies are needed to properly map the spectral index over the western edge of the tail to identify whether gentle re-energization is indeed powering the diffuse emission from the tail. This connection between head-tail radio galaxies and halo emission has been observed in a few other clusters as well \citep[e.g.,][]{2018ApJ...852...65R,2019A&A...622A..22M}.
We also identify a sharp front in the halo, annotated in Figure \ref{fig:radioradioA1132}. This could be indicating a shock or shear motions in the ICM, although it is not visible in the X-ray image presented in \citet{Wilber_2017}. It might also be a magnetic filament or a region of higher turbulence seen in projection. Filamentary emission has been identified in halos before \citep[e.g., in Abell 2255;][]{2005A&A...430L...5G,2020arXiv200604808B}. 
To investigate the possible polarization of the filament, deep higher frequency observations are required.

\section{Conclusion}\label{sec:conclusion}
This study presented a search for diffuse emission in the deepest LOFAR 144 MHz observations ever taken. All Planck Sunyaev-Zel'dovich detected clusters \citep[PSZ2; ComPRASS;][]{2016A&A...594A..27P,2019A&A...626A...7T} and clusters from the Meta Catalogue of X-ray detected Clusters \citep[MCXC;][]{2011A&A...534A.109P} that overlap with the Deep Fields were inspected. The halos were systematically fitted with spherically symmetrical exponential profiles using Markov Chain Monte Carlo sampling to sample the likelihood function. 

We have found a new radio halo in the low mass,  high-redshift cluster MCXCJ1036.1+5713 ($z=0.77$) and tentatively detect diffuse emission from the low mass cluster PSZ2G084.69+42.28 ($z=0.13$). We have set deep upper limits on diffuse emission from clusters with a non-detection and for two clusters previously observed with LOFAR, PSZ2G147.88+53.24 and PSZ2G149.22+54.18, we confirm results in the literature. 

This study has detected diffuse emission in a largely unexplored region of parameter space for galaxy clusters. The results were compared to the radio luminosity - cluster mass relation for radio halos found in the literature, and we found 
that this small sample of clusters is consistent with the correlation extrapolated to lower masses.

The results presented here underline the importance of deep low-frequency observations of galaxy clusters. As the LOFAR Deep Fields reach their final depths of 10-15 $\mu$\Jyb, we expect more low-mass clusters to show radio halos and to put more stringent upper limits on the radio luminosity of lower mass clusters, which will begin to allow a statistical study of a sample of radio halos in low mass clusters.

In the future, international baseline data will additionally be imaged, resulting in sub-arcsecond resolution images at the same depth, allowing better separation of AGN and diffuse emission, especially for mini-halos and high-redshift clusters.

\begin{acknowledgements}
    We thank W. Williams for her plotting functions. 
    EO, RJvW and AB acknowledge support from the VIDI research programme with project number 639.042.729, which is financed by the Netherlands Organisation for Scientific Research (NWO).
    GB, RC, FG, MR acknowledge support from INAF through mainstream program `galaxy clusters science with LOFAR' 1.05.01.86.05.
    Ann.B. acknowledges support from the ERC-Stg DRANOEL n. 714245 and from the MIUR FARE grant “SMS”.
    PNB is grateful for support from the UK STFC via grant ST/R000972/1. 
    MBo acknowledges support from INAF under PRIN SKA/CTA FORECaST and from the Ministero degli Affari Esteri della Cooperazione Internazionale - Direzione Generale per la Promozione del Sistema Paese Progetto di Grande Rilevanza ZA18GR02. 
    GDG acknowledges support from the ERC Starting Grant ClusterWeb 804208. 
    MJH acknowledges support from the UK Science and Technology Facilities Council (ST/R000905/1).
    HR acknowledges support from the ERC Advanced Investigator programme NewClusters 321271.
    JS is grateful for support from the UK STFC via grant ST/R000972/1.

    LOFAR \citep{Haarlem2013} is the Low Frequency Array designed and constructed by
    ASTRON. It has observing, data processing, and data storage facilities in several countries,
    which are owned by various parties (each with their own funding sources), and that are
    collectively operated by the ILT foundation under a joint scientific policy. The ILT resources
    have benefited from the following recent major funding sources: CNRS-INSU, Observatoire de
    Paris and Université d'Orléans, France; BMBF, MIWF-NRW, MPG, Germany; Science
    Foundation Ireland (SFI), Department of Business, Enterprise and Innovation (DBEI), Ireland;
    NWO, The Netherlands; The Science and Technology Facilities Council, UK; Ministry of
    Science and Higher Education, Poland; The Istituto Nazionale di Astrofisica (INAF), Italy.
    This research made use of the Dutch national e-infrastructure with support of the SURF
    Cooperative (e-infra 180169) and the LOFAR e-infra group. The Jülich LOFAR Long Term
    Archive and the German LOFAR network are both coordinated and operated by the Jülich
    Supercomputing Centre (JSC), and computing resources on the supercomputer JUWELS at JSC
    were provided by the Gauss Centre for Supercomputing e.V. (grant CHTB00) through the John
    von Neumann Institute for Computing (NIC).
    This research made use of the University of Hertfordshire high-performance computing facility
    and the LOFAR-UK computing facility located at the University of Hertfordshire and supported
    by STFC [ST/P000096/1], and of the Italian LOFAR IT computing infrastructure supported and
    operated by INAF, and by the Physics Department of Turin university (under an agreement with
    Consorzio Interuniversitario per la Fisica Spaziale) at the C3S Supercomputing Centre, Italy.
    This research has made use of NASA's Astrophysics Data System.
    
\end{acknowledgements}

\bibliographystyle{aa}
\bibliography{firstbib.bib}

\clearpage
\section*{Appendix I - Surface Brightness Fits}
\begin{figure}[h!]
    \centering
    \includegraphics[width=1.0\textwidth]{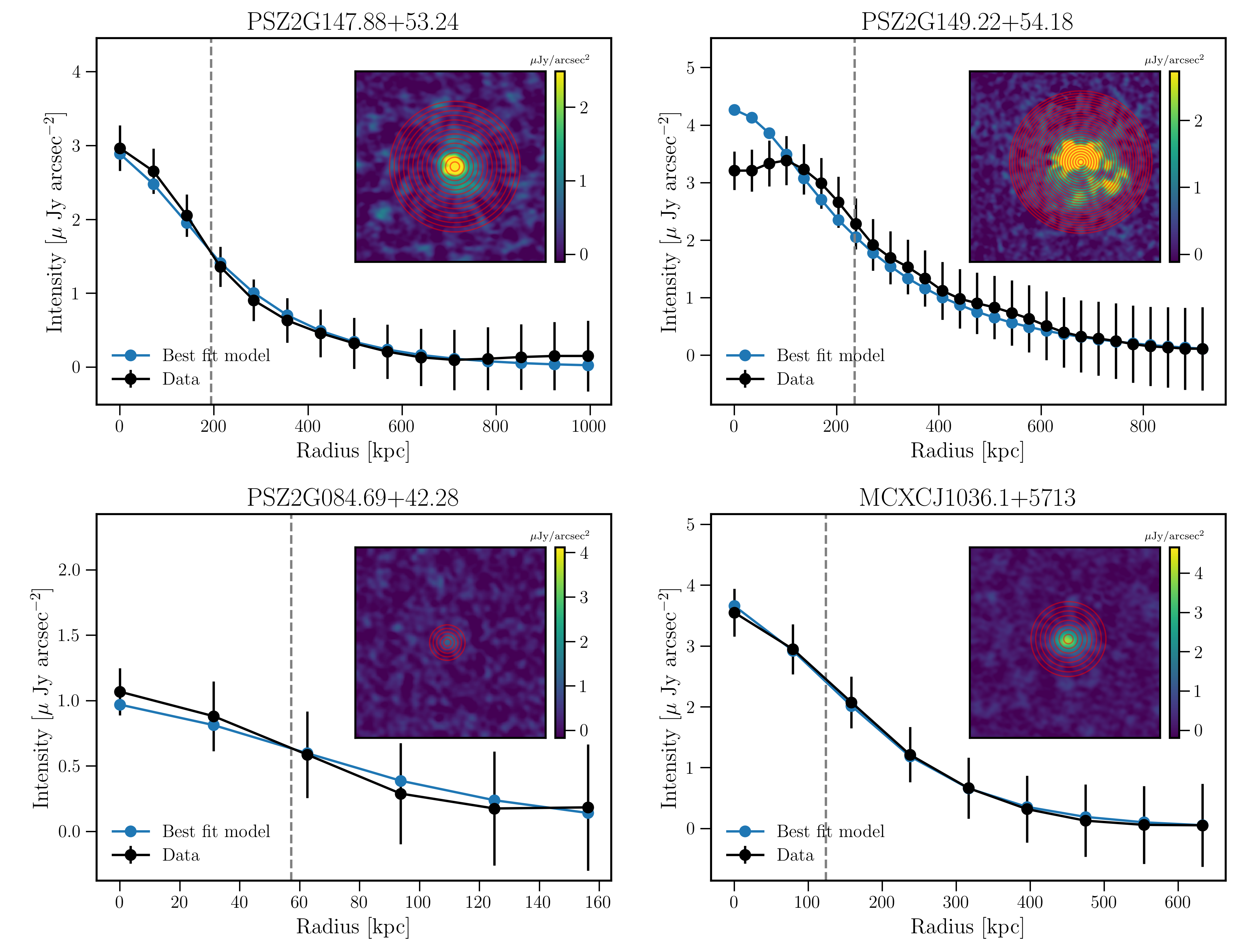}
    \caption{Azimuthally averaged surface brightness profiles for the clusters presented in this work. The inset images show the concentric annuli where the profile has been calculated, the width of the annuli is equal to the semi-major axis of the restoring beam. The dashed gray line indicates the best-fit $e$-folding radius.}
    \label{fig:radial}
\end{figure}
\clearpage
\begin{figure}[h!]
    \centering
    \includegraphics[width=0.8\textwidth]{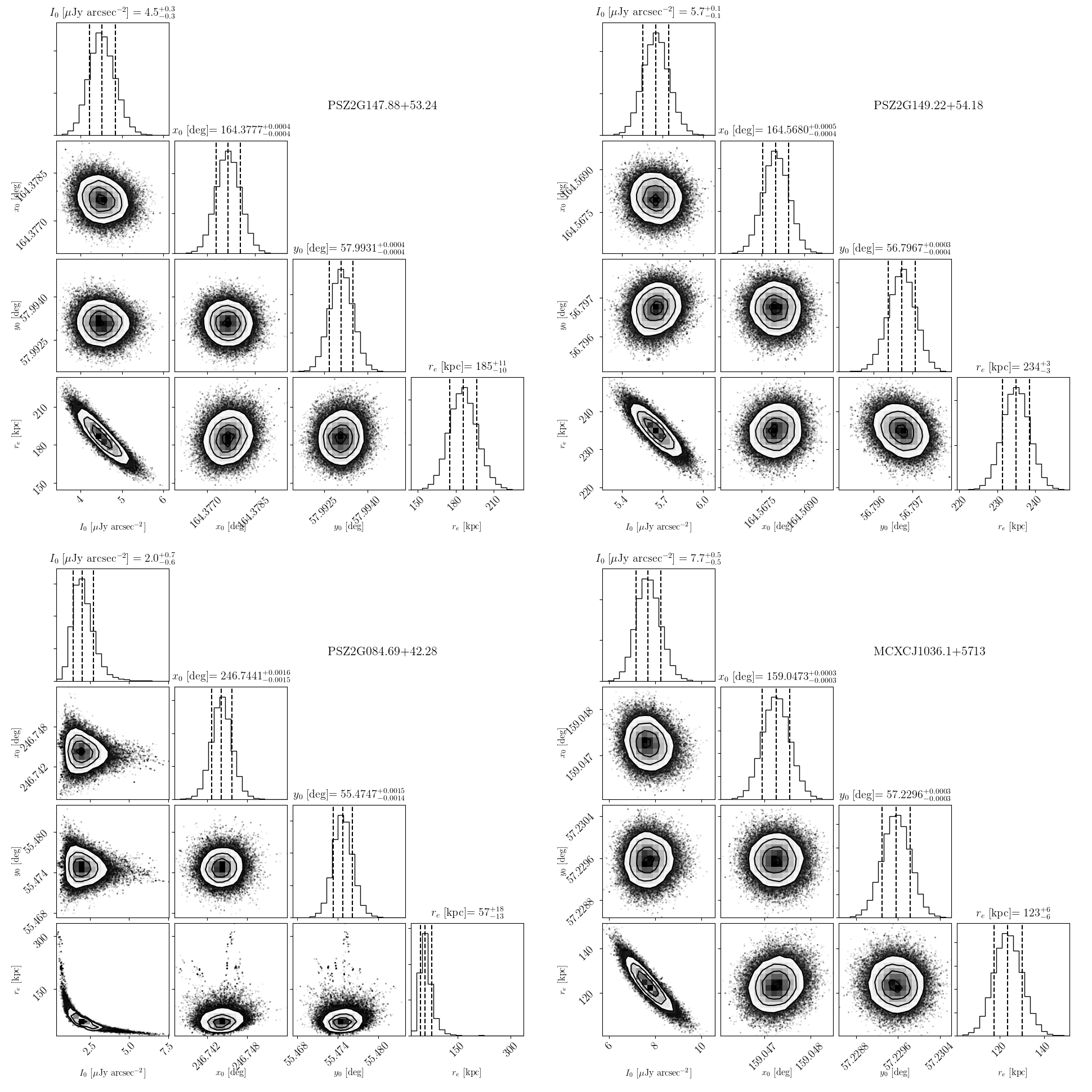}
    \caption{Corner plots of the fitted parameters given by the MCMC chain. The dashed lines indicate the 16, 50th and 84th percentiles of the chain and the solid line indicates the initial value found by a least-square fit.}
    \label{fig:my_label}
\end{figure}

\clearpage

\section*{Appendix II - Radio Optical overlays}

\begin{figure}[h!]
    \centering
    \includegraphics[width=1.0\columnwidth]{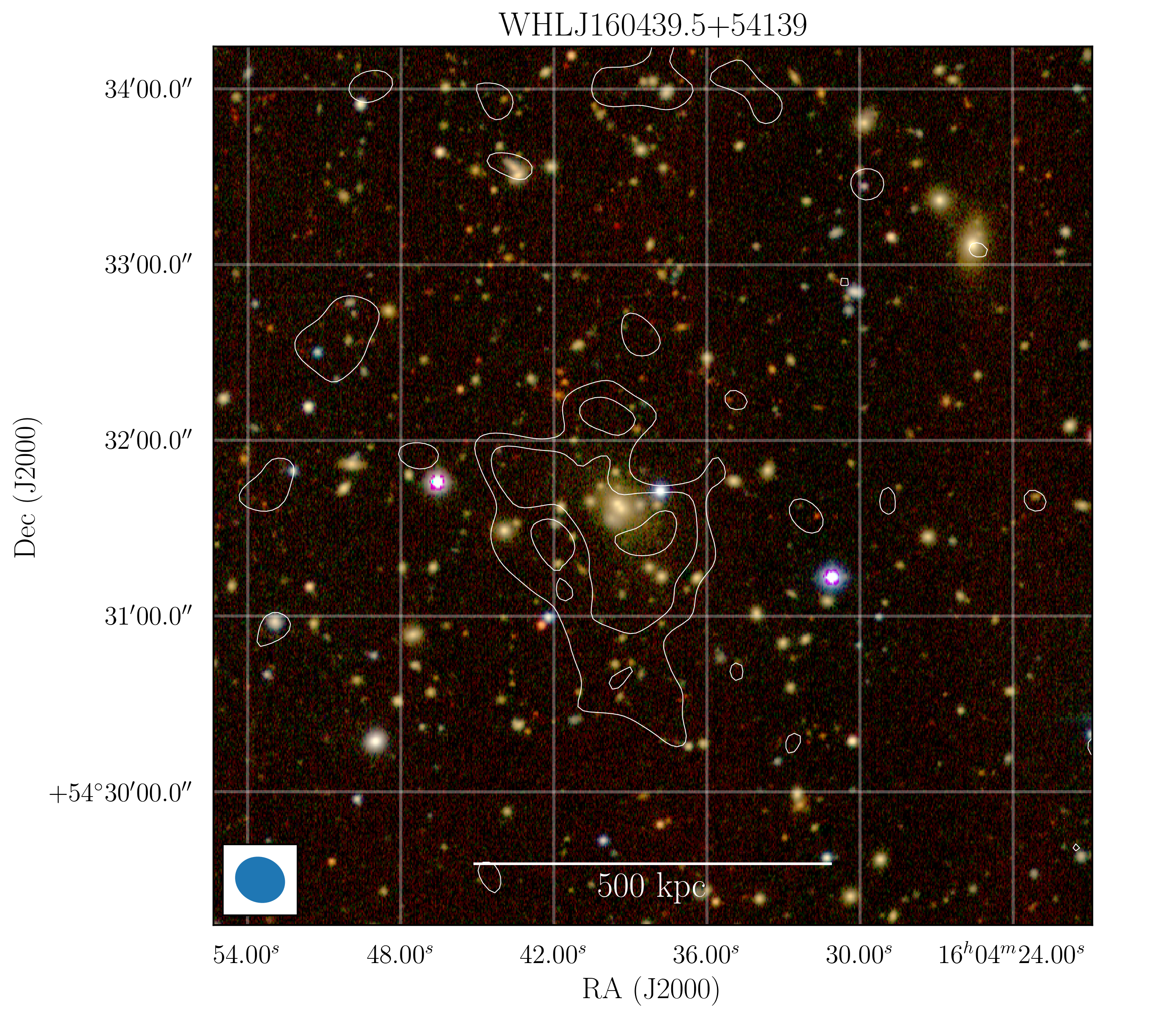}
    \caption{Optical ($grz$) image of WHLJ160439.5+54139 from the Legacy Survey with compact source subtracted low-resolution LOFAR contours overlaid. The beam size is $17''\times15''$. Contours at $[3,6,12,..]\sigma$, where $\sigma = 72\mu$\Jyb.}
    \label{fig:WHL1optical}
\end{figure}

\begin{figure}[h!]
    \centering
    \includegraphics[width=1.0\columnwidth]{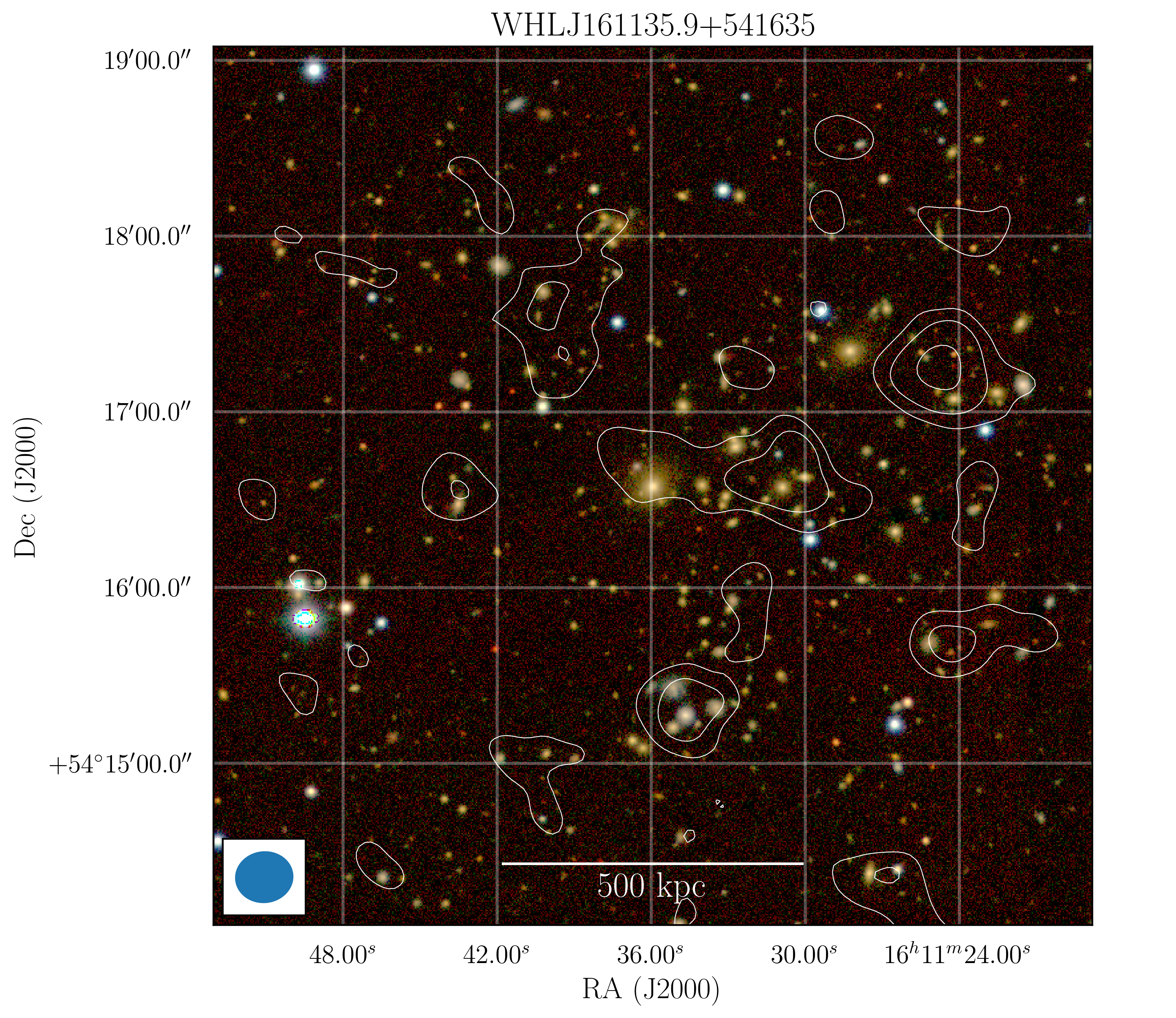}
    \caption{Optical ($grz$) image of WHLJ161135.9+541635 from the Legacy Survey with compact source subtracted low-resolution LOFAR contours overlaid. The beam size is $20''\times 18''$. Contours at $[3,6,12,..]\sigma$, where $\sigma = 62\mu$\Jyb.}
    \label{fig:WHL2optical}
\end{figure}

\begin{figure}[h!]
    \centering
    \includegraphics[width=1.0\columnwidth]{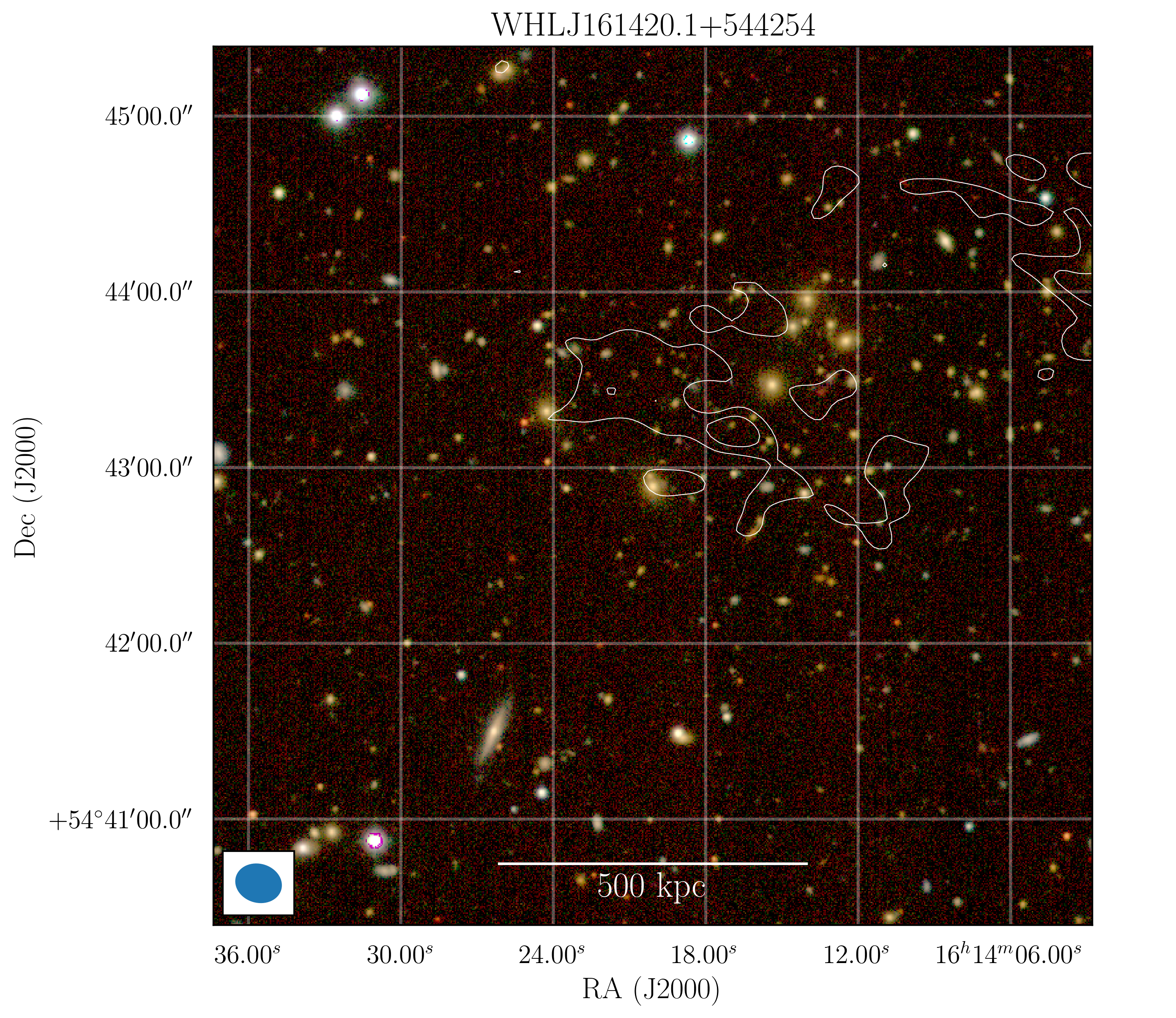}
    \caption{Optical ($grz$) image of WHLJ161420.1+544254 from the Legacy Survey with compact source subtracted low-resolution LOFAR contours overlaid. The beam size is $X16\times 13''$. Contours at $[3,6,12,..]\sigma$, where $\sigma = 66\mu$\Jyb.}
    \label{fig:WHL3optical}
\end{figure}

\begin{figure}[h!]
    \centering
    \includegraphics[width=1.0\columnwidth]{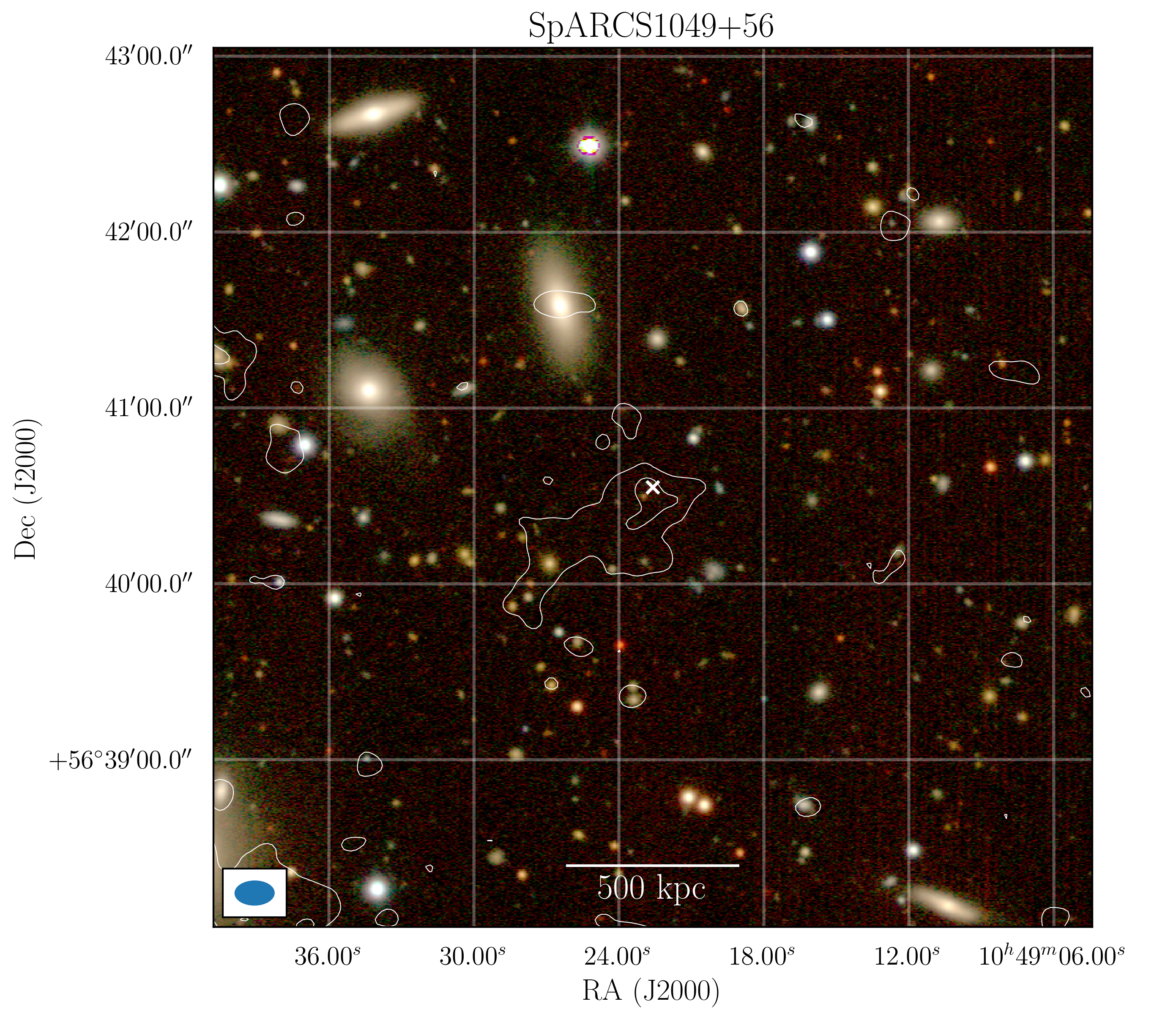}
    \caption{Optical ($grz$) image of SpARCS1049+56 from the Legacy Survey with compact source subtracted low-resolution LOFAR contours overlaid. The beam size is $14''\times 9''$. Contours at $[3,6]\sigma$, where $\sigma = 48\mu$\Jyb.}
    \label{fig:SPARCSoptical}
\end{figure}

\begin{figure}
    \centering
    \includegraphics[width=1.0\columnwidth]{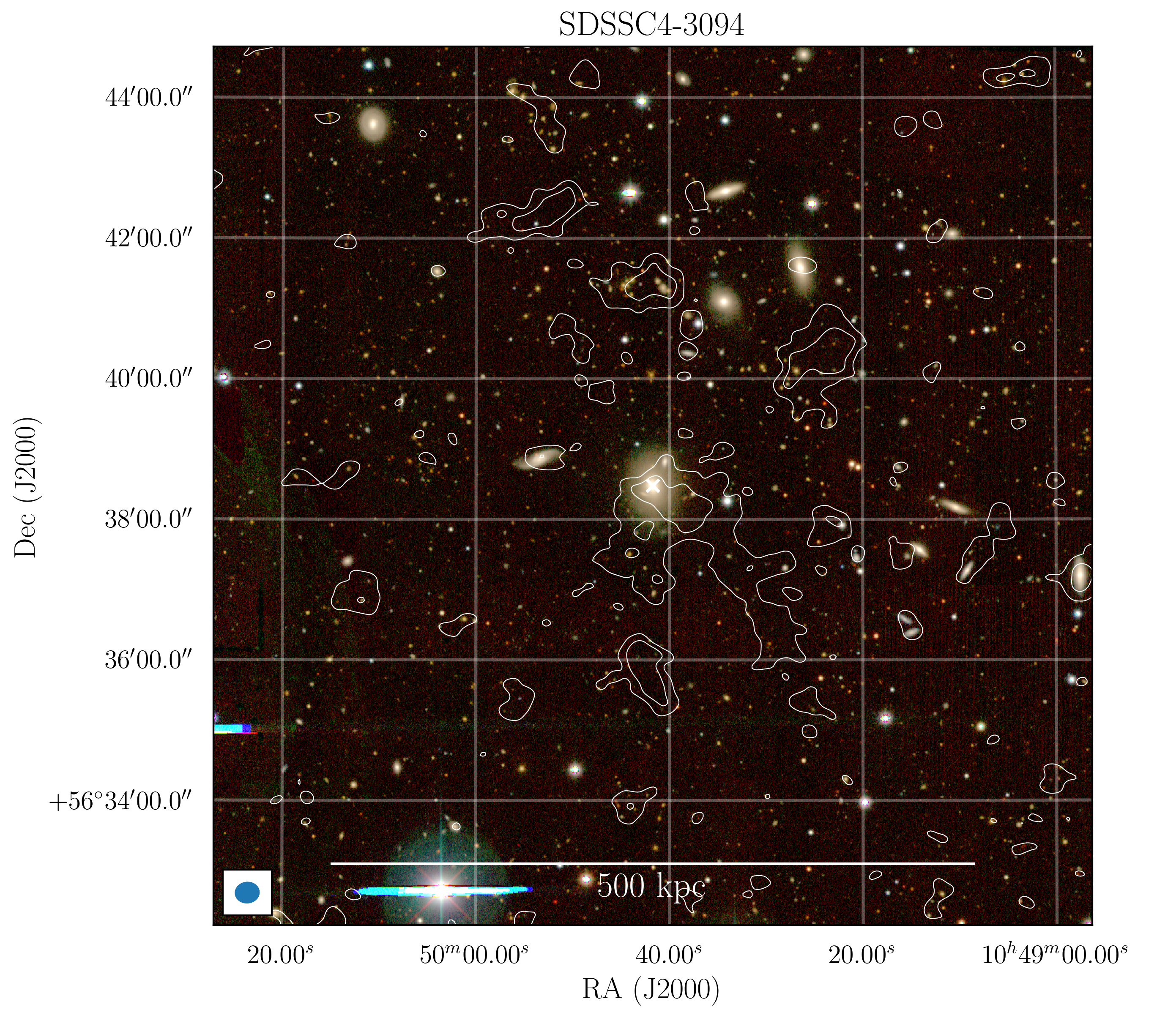}
    \caption{Optical ($grz$) image of SDSSC4-3094 from the Legacy Survey with compact source subtracted low-resolution LOFAR contours overlaid. The beam size is $21''\times 19''$. Contours at $[3,6,12,..]\sigma$, where $\sigma = 74\mu$\Jyb.}
    \label{fig:SDSSoptical}
\end{figure}


\end{document}